\documentclass[a4paper,11pt]{article}
\pdfoutput=1 

\usepackage{jcappub} 

\usepackage[T1]{fontenc} 
\usepackage{graphicx} 
\usepackage{subcaption}
\usepackage{xcolor} 
\usepackage{ulem} 
\usepackage{braket}
\usepackage{hyperref}
\usepackage{multirow}
\usepackage{amsmath} 
\usepackage{subcaption} 
\usepackage{rotating}

\title{Taking the Weight Off: Mitigating Parameter Bias from Catastrophic Outliers in 3$\times$2pt Analysis}

\author[a]{Carolyn McDonald Mill}
\author[a]{, C. Danielle Leonard}
\author[a]{, Markus Michael Rau}
\author[b]{, Cora Uhlemann}
\author[c]{, and Shahab Joudaki}

\affiliation[a]{School of Mathematics, Statistics and Physics, Newcastle University, Herschel Building, NE1 7RU Newcastle-upon-Tyne, U.K.}
\affiliation[b]{Fakultät für Physik, Universität Bielefeld, Postfach 100131, 33501 Bielefeld, Germany}
\affiliation[c]{ Centro de Investigaciones Energéticas, Medioambientales y Tecnológicas (CIEMAT), Av. Complutense 40, E-28040 Madrid, Spain}

\emailAdd{c.mill2@newcastle.ac.uk}
\emailAdd{danielle.leonard@newcastle.ac.uk}
\emailAdd{markus.rau@newcastle.ac.uk}
\emailAdd{cuhlemann@physik.uni-bielefeld.de}
\emailAdd{shahab.joudaki@ciemat.es}

\abstract{
Stage IV cosmological surveys will map the universe with unprecedented precision, reducing statistical uncertainties to levels where unmodelled systematics can significantly bias inference. In particular, photometric redshift (photo-$z$) errors and intrinsic alignments (IA) must be robustly accounted for to ensure accurate inference of cosmological parameters. The increasing depth of Stage IV surveys exacerbates these challenges by producing low signal-to-noise galaxy populations prone to inaccurate photo-$z$ measurements. Catastrophically misidentified redshifts are especially problematic for 3$\times$2pt inferences that combine weak lensing and galaxy clustering information. We demonstrate that even modest outlier fractions (e.g. 5\%) can lead to substantial biases in cosmological parameter estimates: up to 1.8$\sigma$ in $\Omega_M$ and $\sigma_8$, and over 8$\sigma$ in the IA redshift evolution parameter $\eta$. To address this, we introduce a flexible weighting scheme at the likelihood level that down-weights the most contamination-sensitive elements of the data vector during inference. This method mitigates biases without inflating the parameter space, reducing cosmological parameter biases to below 1$\sigma$ without substantially degrading constraining power. Our approach offers a practical solution for future analyses, enabling robust cosmological inference in the presence of catastrophic redshift errors.}

\begin{document}
\newcommand{\comment}[1]{\textcolor{red}{#1}}
\newcommand{\cu}[1]{\textcolor{blue}{#1}}
\newcommand{\dl}[1]{\textcolor{teal}{#1}}

\maketitle
\flushbottom

\section{Introduction}

Upcoming and ongoing Stage IV surveys, such as the Vera C. Rubin Observatory Legacy Survey of Space and Time (LSST) \citep{Ivezi__2019}, \textit{Euclid} \citep{laureijs2011eucliddefinitionstudyreport,2025EuclidOverview}, and the Nancy Grace Roman Space Telescope \citep{spergel2015widefieldinfrarredsurveytelescopeastrophysics}, will offer an unprecedented scale and precision of late-time cosmological data. These surveys promise to help us answer pressing cosmological questions such as the nature of dark energy \citep{divalentino2025cosmoversewhitepaperaddressing,2025EuclidOverview,lsstdarkenergysciencecollaboration2012largesynopticsurveytelescope,merloni2012erositasciencebookmapping}. 

Weak lensing \citep{Bartelmann_2001,Mandelbaum_2018} is a powerful probe of cosmological parameters, particularly those which characterise dark energy \citep{Huterer_2010,Abbott_2023,Joudaki_2017}. Typical analyses often detect this effect via two-point correlations of the shapes of galaxies lensed by the matter distribution between them and the observer. This allows us to extract information about the intervening matter along our line of sight to them - a signal known as cosmic shear. Cosmic shear can be complemented with two-point correlations capturing galaxy clustering and galaxy-galaxy lensing - the cross-correlation between shapes and positions - to produce a combined set of two-point correlations, often referred to as 3$\times$2pt statistics. 

However, as the precision of our data improves, the dominant source of error shifts from statistical uncertainties to systematic errors, i.e., those arising from our modelling \citep{Huterer_2006, zhang2025forecastingimpactsourcegalaxy, Sugiyama_2023, Secco_2022, kids-450}.  It is therefore essential to adequately include systematic effects in our models and understand how they may impact our analysis, ensuring that this influx of high-quality data can be fully exploited.

In order to probe the nature of dark energy, we require information on how the large-scale structure has evolved over time. Galaxy redshifts are essential for tracking this evolution. Given the billions of galaxies targeted by surveys like LSST and the corresponding hypothetical cost of obtaining spectra for all galaxies, redshifts for such surveys are estimated primarily through photometric methods - observing galaxies in a set of broad bands and applying redshift estimation techniques such as template fitting and machine learning \cite{salvato2018flavoursphotometricredshifts}. While photometric redshifts (photo-$z$) require less exposure time and are less computationally intensive to obtain than spectroscopic redshifts (spec-$z$), they are also significantly less accurate \cite{Newman_2022}. To account for photo-$z$ uncertainties in tomographic analyses, nuisance parameters that characterise uncertainties in galaxy redshift distributions are introduced in the data model. Often they are parametrised as small shifts in the mean and width of redshift distributions, and their value is inferred alongside cosmological parameters \citep{ Hildebrandt_2020, Sugiyama_2023}. However, it remains unclear whether this approach sufficiently captures the whole range of ways in which photo-$z$ errors qualitatively impact source galaxy redshift distributions. 

One type of photo-$z$ error which is generally un-modelled in state-of-the-art cosmological analyses is the presence of catastrophic outliers: redshift estimates whose residuals are several times larger than the typical photo-$z$ scatter, placing the galaxy in the wrong tomographic bin or at a vastly different cosmic epoch. Multiple photometric redshifts are often plausible for a given photometry, as demonstrated in \cite{2019MNRAS.489..820B}. Such outliers commonly originate from degeneracies in spectral energy distribution (SED) fitting, where distinct spectral breaks produce similar broadband colours -- for instance, confusion between the Lyman break at high redshift and the Balmer break at low redshift \citep{Jouvel_2011}. Failing to account for catastrophic outliers has been shown to bias our inference of cosmological parameters \citep{Hearin_2010,Bernstein_2010}. These outlier populations are difficult to model because of the scarcity of spectroscopic redshifts at the depths probed by Stage IV surveys. This makes it challenging to construct representative training samples to reconstruct these populations \citep{Masters_2015}.

In addition to photo-$z$ systematics, weak lensing measurements are also impacted by intrinsic galaxy alignments (IA) \citep{Troxel_2015,Joachimi_2015, Lamman_2024}. The observed cosmic shear signal consists of correlations in both the distortion of galaxy images due to the cosmic structure and the intrinsic galaxy shapes. Correlated intrinsic shapes, caused by physical effects such as tidal forces, can contaminate the weak lensing signal. Understanding and correctly modelling intrinsic alignments is thus crucial for extracting accurate cosmological information, as incorrect parametrisation of this contribution has been shown to be susceptible to biasing cosmological parameters \citep{Krause_2015,Kirk_2015}.

Previous studies have found that significant bias in cosmological parameters can be induced by misspecifying the modelling of either IA \citep{Krause_2015,Kirk_2015,Yao_2017} or photo-$z$ uncertainty \citep{EuclidXXXI_2024,awan2024impactlargescalestructuresystematics,wright2025kidslegacycosmologicalconstraintscosmic}, where misspecification refers to adopting an analysis model that omits or inaccurately represents relevant physical observational effects. Moreover, a coupling between these systematics has been identified \citep{wright2020kids, stolzner2021self}, and it has been demonstrated that the degeneracies between their parametrisations can further exacerbate these biases \citep{Leonard_2024,Fischbacher_2023}.

In this work, we investigate the sensitivity of cosmological parameter inference to the misspecification of photo-$z$ and IA modelling. We perform a Markov Chain Monte Carlo (MCMC) analysis on a synthetic 3$\times$2pt LSST-like data set, implementing parametrisations for IA and photo-$z$, along with a model for an outlier population arising from the confusion between Lyman and Balmer breaks.  We examine the impact of misspecifying this outlier population on our cosmological and IA parameters and investigate the potential of modifying the likelihood as a method to reduce this effect.

We describe the models used for inference and for quantifying model misspecification in Section \ref{section:modelling}. In Section \ref{section:results} we describe the numerical choices made for this analysis and present our results, focusing on constraints on cosmological parameters $\Omega_M, \sigma_8$, dark energy equation of state parameters $w_0, w_a$, and IA parameters for the non-linear alignment model $a, \eta$. We discuss our findings and outline our conclusions in Section \ref{section:discuss}. 

\section{Set-up and Modelling}
\label{section:modelling}

\begin{figure}
    \centering
    \includegraphics[width=\linewidth]{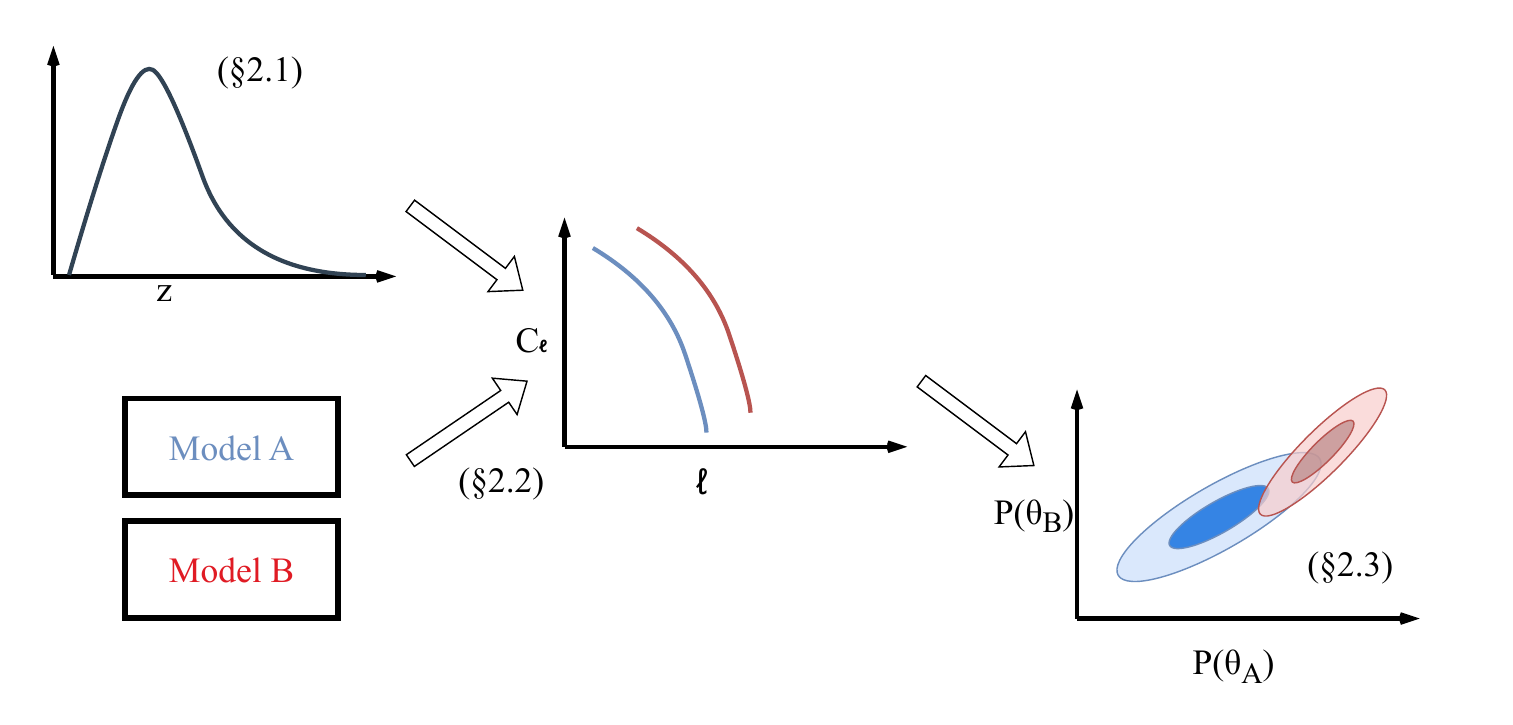}
    \caption{The cosmological inference process from a galaxy sample over redshift range $z$, to summary statistic $C_\ell$, to parameters of interest $\theta_i$.}
    \label{fig:overview_2pt}
\end{figure}

In Figure \ref{fig:overview_2pt} we sketch the inference process for determining cosmological parameter values; we will provide details of how each step was carried out for this work in the indicated subsections. From a population of galaxies described by a photo-$z$ distribution (\ref{section:photoz}) we obtain summary statistics (in our case the angular correlation functions, $C_\ell$) that we predict from a model with a chosen parameter set (\ref{section:summmarystat}). We can then use these statistics to infer information about cosmological parameters from the data (\ref{section:inference}). However, it is possible that our model does not perfectly represent the data, which can lead to biases in the inferred values of parameters $\theta_i$. In this Section, we also discuss our method to test how well a given model defined by a set of parameters represents the data. We then introduce composite likelihoods as a method of improving model performance under misspecification. 

\subsection{Photometric Redshifts}
\label{section:photoz}

\begin{figure}[htbp]
\centering
\includegraphics[width=.45\textwidth]{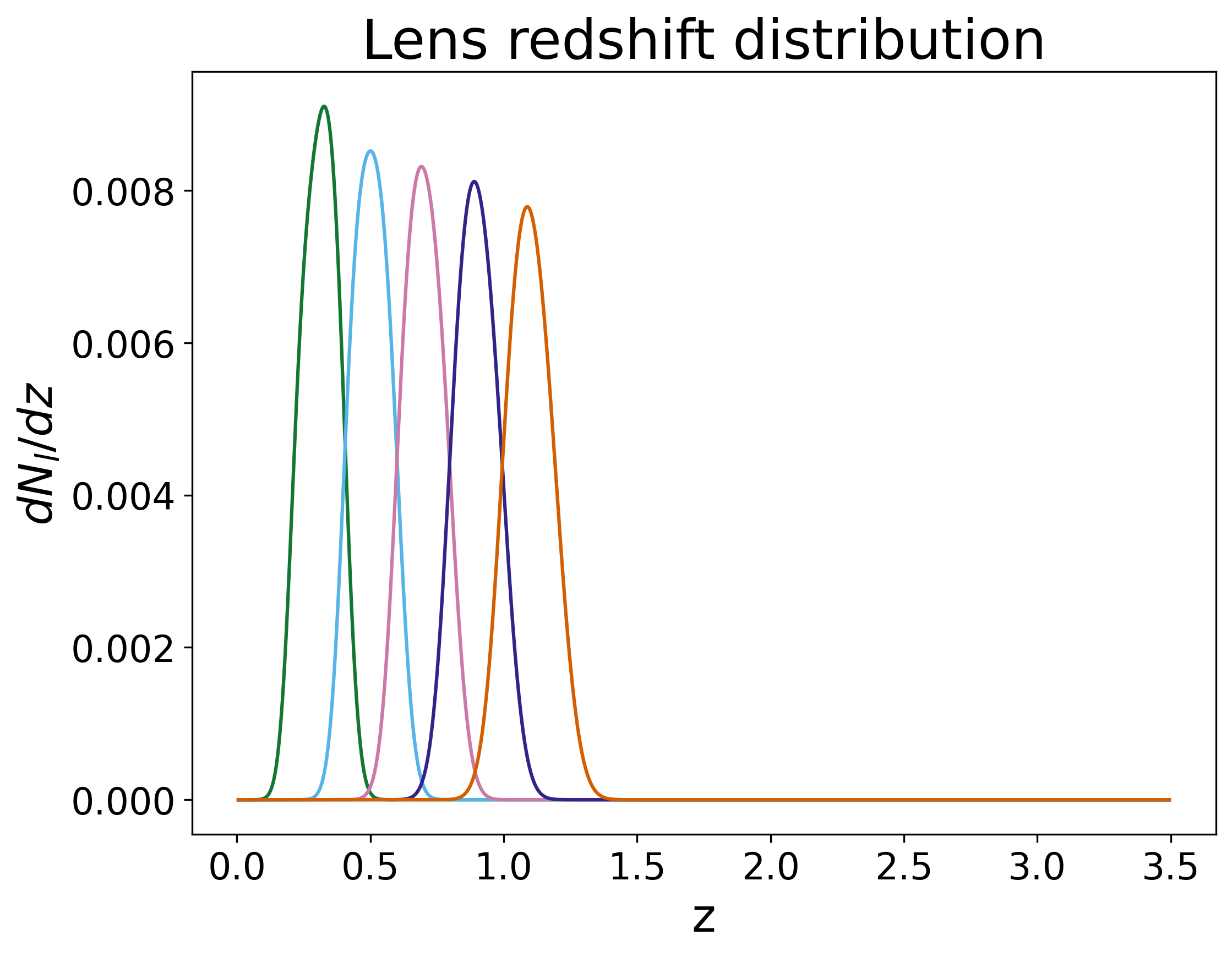}
\qquad
\includegraphics[width=.45\textwidth]{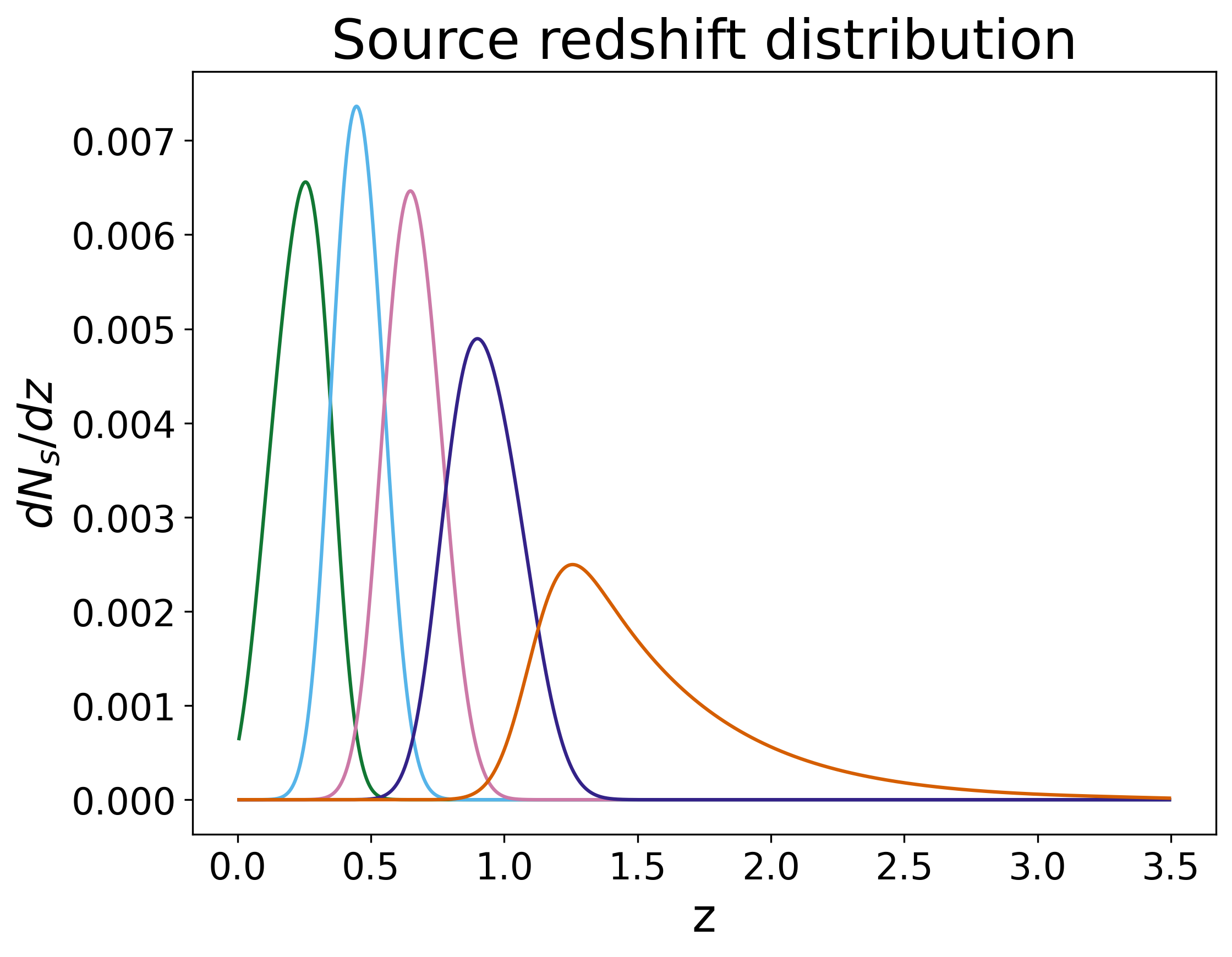}
\caption{Redshift distributions for the five bins of the lens (left) and source (right) samples.}\label{fig:gal_dist}
\end{figure}

Photo-$z$ methods allow redshift information to be obtained with less exposure time and more cost-effectively than with spec-$z$ methods across the vast number of galaxies observed in large-scale sky surveys. In cosmological analysis, galaxies with similar redshifts are typically grouped into tomographic redshift bins, enabling us to probe the statistical behaviour of structures at different epochs rather than on a per-galaxy basis. In this work, we use mock redshift distributions for the lens and source samples (used for clustering and cosmic shear analyses respectively, with both in use for galaxy-galaxy lensing), each convolved with Gaussian distributions to represent a Gaussian model for photo-$z$ mean shift and variance uncertainty. Throughout this work, tomographic bins are indexed from 0 to 4 (rather than the conventional 1 to 5) for consistency with our analysis pipeline, with 0 being the lowest-redshift bin. 

{\it Mean shifts.} Small errors in estimated galaxy redshifts can arise from noise in the photometric measurements, limitations of the template or training set used for estimation, and mild degeneracies between spectral features. To account for these uncertainties in inference, cosmological analyses often introduce nuisance parameters that allow small shifts in the mean of each redshift bin to be marginalised over \citep{Hoyle_2018,Hildebrandt_2020,Sugiyama_2023}. In this analysis, we model photo-$z$ uncertainty using two sets of five nuisance parameters, $\Delta z^i$, which allow for independent shifts in the mean redshift of each tomographic lens and source bin. These parameters modify the redshift bins such that 
\begin{equation}
    z_{new}^{i} = z^i + \Delta z^i.
\end{equation}

{\it Catastrophic Outliers.}
Catastrophic outliers concern any case where the photometric redshift distribution diverges significantly from the true redshift distribution of the sample. This can be caused by the blending of samples or the redshifting of distinctive spectroscopic features outside of the photometric broadbands \cite{2019MNRAS.489..820B}. 

A dominant case of outliers is caused by the degeneracy between high-$z$ Lyman-break galaxies and low-$z$ Balmer break galaxies, producing a contaminating population of galaxies that are determined to have redshift values $z\sim0-0.5$  when their true redshift will lie around $z\sim2.5-3.5$ \cite{Graham_2017}. 

In the case of LSST, the source sample is expected to have a more prominent presence of outliers due to the less stringent requirements on the photo-$z$ uncertainty in these objects relative to the lenses.

\begin{figure}
    \centering 
    \includegraphics[width=0.6\textwidth]{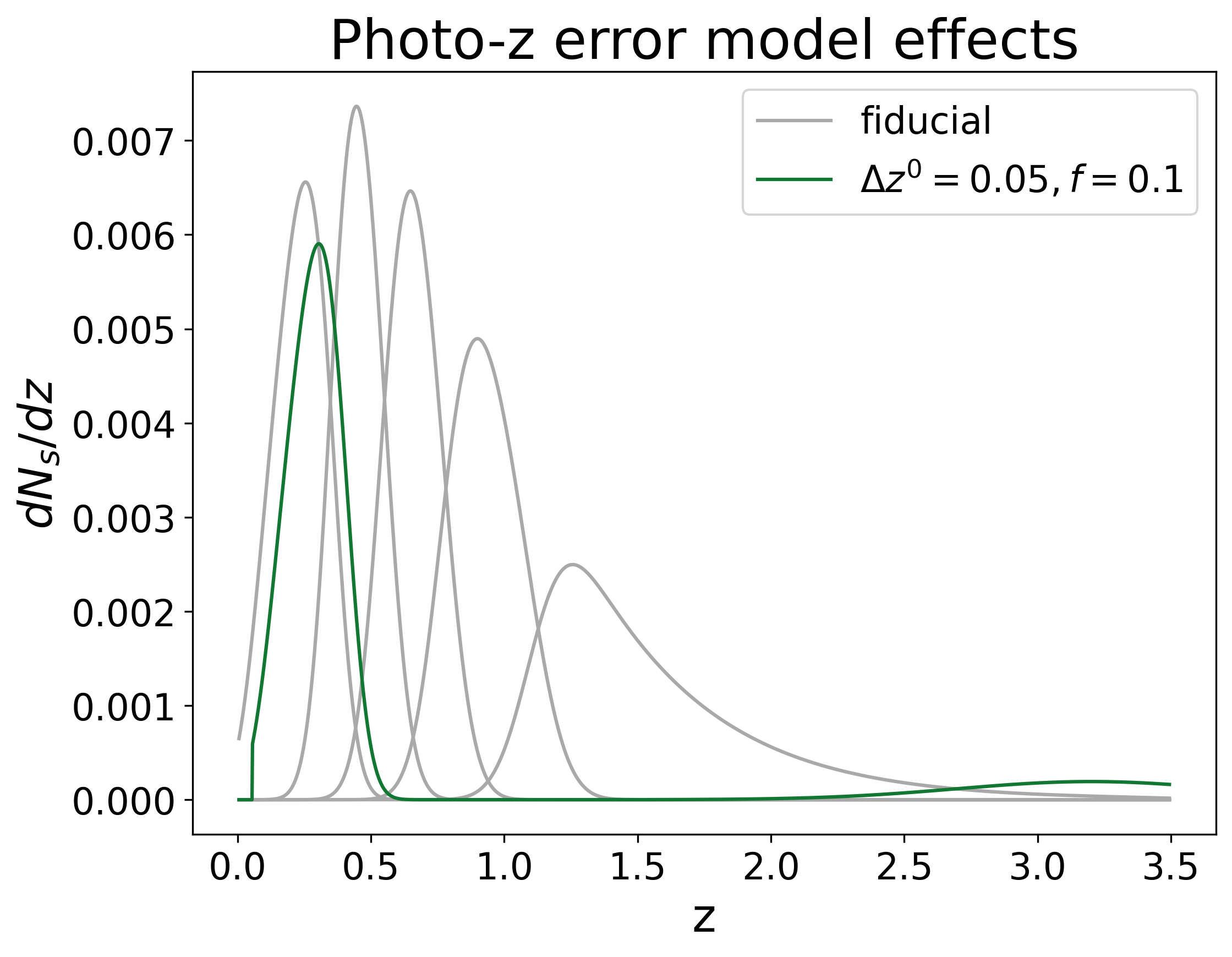}
    \caption{The photo-$z$ distribution of source bin $\frac{dN_s^0}{dz}$ with mean shift $\Delta z^0 = 0.05$ and an outlier fraction of $f=0.1$. We use unrealistically large values here for clear visualisation. The fiducial source bins are plotted in grey for comparison.}
    \label{fig:outlier_dNdz}
\end{figure}

We examine the effects of outliers on parameter inference by including the effect of a Lyman-Balmer confusion in the mock data. We do this by allowing for a fraction $f$ of the total photo-$z$ distribution of $\frac{dN_s^0}{dz}$ (the redshift distribution of source bin 0) to be an outlier population with a Gaussian redshift distribution centred at $z=3.2$ with a width of $0.5$, such that the total distribution of the photo-$z$ in source bin 0 is given by
\begin{equation}
 \frac{d N_s^0(f)}{dz}=   (1-f) \frac{dN_s^{0,fid}}{dz} + f \mathcal{N}(3.2,0.5)
\end{equation}
This distribution is chosen to approximate the prominent case of outliers due to the confusion between Lyman break and Balmer break galaxies \citep{Graham_2017}. We demonstrate the joint effect that a mean shift $\Delta z^0=0.05$ and an outlier fraction of $f=0.1$ has on the redshift distribution as the green line in Figure~\ref{fig:outlier_dNdz}. 

\subsection{3$\times$2pt}
\label{section:summmarystat}

3$\times$2pt is the expected baseline approach for analysing weak lensing and photometric galaxy clustering in Stage IV surveys. It combines the constraining power from two-point statistics of galaxy clustering, galaxy-galaxy lensing, and weak lensing using their two-point angular power spectra. The total datavector in our model is constructed using the 5 autocorrelations between the galaxy clustering redshift bins, 7 galaxy-galaxy lensing redshift bin combinations\footnote{[0,2], [0,3], [0,4], [1,3], [1,4], [2,4], [3,4], taken from \citep{thelsstdarkenergysciencecollaboration2021lsstdarkenergyscience}} and the 15 unique cosmic shear auto and cross redshift bin combinations. 

The two-point angular auto-power spectra for galaxy clustering can be formulated as
\begin{equation}
    C_{g^i g^i}^\ell = b_i^2 \int_0^{\chi_\infty} d\chi \left(\frac{dN_l^i}{d\chi}\right)^2 P_{mm}\left(\frac{\ell}{\chi},\chi\right)\,,
\end{equation} 
where $b_i$ denotes the linear galaxy bias, $\frac{dN^i}{d\chi}$ is the photometric redshift distribution in photo-$z$ bin $i$ with respect to comoving distance $\chi$, $\chi_\infty$ is the comoving distance at the horizon, and $P_{mm}$ is the non-linear matter power spectrum. We assume a flat universe and adopt the Limber approximation throughout. 

The galaxy-galaxy lensing angular power spectra are given by
\begin{equation}
    C_{\kappa^i g^j}^\ell = b_j \int_0^{\chi_\infty} d\chi \frac{W_i(\chi)}{\chi} \frac{dN_l^j}{d\chi}P_{m m}\left(\frac{\ell}{\chi},\chi\right)\,,
\end{equation}
where the lensing efficiency kernel $W_i(\chi)$ defined as
\begin{equation}
    W_i(\chi) = \frac{3 H_0^2 \Omega_M}{2c^2} \frac{\chi}{a(\chi)}\int_\chi^{\chi_\infty}\frac{\chi'-\chi}{\chi'} \frac{dN_s^i}{d\chi'}d\chi' \,.
\end{equation}
Finally, the weak lensing angular power spectra are given by 
\begin{equation}
    C_{\kappa^i \kappa^j}^\ell=\int_0^{\chi_\infty} d\chi \frac{W_i(\chi) W_j(\chi)}{\chi^2}P_{m m}\left(\frac{\ell}{\chi}, \chi \right)\,.
\end{equation}
We computed the angular power spectra for the MCMC using the \texttt{pyccl}\footnote{https://github.com/LSSTDESC/CCL} Python package \cite{Chisari_2019}, where the matter power spectra were calculated using 1-loop  perturbation theory, with calculations carried out using \texttt{FASTPT} \citep{McEwen_2016,Fang_2017}. We assumed a $w_0 w_a$CDM cosmology, parametrised with $\Omega_B$, $\Omega_M$, $n_s$, $\sigma_8$, $h$ and dark energy equation of state parameters $w_0$, $w_a$. 

\textit{Intrinsic Alignments.} The observed projected galaxy shape (or ellipticity) field receives contributions both from weak lensing and from IA. This total effective shear field can be expressed as
\begin{equation}
    \gamma^\epsilon = \gamma^\kappa + \gamma^{IA}\,,
\end{equation}
where we denote the observable shear with $\epsilon$, the shear induced by lensing with $\kappa$ and the intrinsic component due to galaxy shapes with IA. 
Intrinsic alignments arise from the tendency of galaxies to align their shapes with the tidal gravitational field of their surrounding matter environment, together with other environmental and galaxy-evolution related effects. In the widely used Non-Linear Alignment (NLA) model, the IA shear field is assumed to respond linearly to the nonlinear local tidal field tensor \citep{Bridle_2007}. The IA shear is then given by
\begin{equation}
    \gamma_{ij}^{IA} = A_1 s_{ij}\,,
\end{equation}
 where $s_{ij}=\partial_i\partial_j \phi$  is the tidal field tensor derived from the gravitational potential $\phi$. The amplitude of the intrinsic alignment signal is often taken to be a redshift dependent function $A_1(z)$, which captures the strength and redshift evolution of the alignment. Following the formulation in \cite{Secco_2022}, this can be expressed as
\begin{equation}
    A_1(z) = - a \frac{\bar{C}_1 \rho_{c} \Omega_M}{D(z)} \left(\frac{1+z}{1+z_{piv}}\right)^{\eta},
\end{equation}
where $\rho_c$ is the critical density, $D(z)$ is the linear growth factor, the pivot redshift $z_{piv}=0.62$, and the normalisation factor $\bar C_1 = 5 \times 10^{-14} M_\odot h^{-2}\mathrm{Mpc^2}$. 

For cosmic shear, the total angular power spectra for the sum of the cosmic shear $\kappa$ and the intrinsic component $I$ can be written as 
\begin{equation}
    C^\ell_{\epsilon^i \epsilon^j} = C^\ell_{\kappa^i \kappa^j} + C^\ell_{\kappa^i I^j} + C^\ell_{I^i \kappa^j} + C^\ell_{I^i I^j}
\end{equation}
and the galaxy-galaxy lensing case as 
\begin{equation}
    C^\ell_{\epsilon^i g^j} = C^\ell_{\kappa^i g^j} + C^\ell_{I^i g^j}.
\end{equation}
These IA contributions are calculated through line-of-sight projections of their respective 3D power spectra
\begin{equation}
    C_{\kappa^i I^j}^\ell = \int_0^{\chi_\infty} d\chi \frac{W_i(\chi)}{\chi} \frac{dN_s^j}{d\chi} P_{mI}\left(\frac{\ell}{\chi},\chi\right),
\end{equation}
\begin{equation}
        C_{I^i I^j} = \int_0^{\chi_\infty} \frac{1}{\chi^2} \frac{dN_s^i}{d\chi}\frac{dN_s^j}{d\chi} P_{I I}\left(\frac{\ell}{\chi},\chi\right),
\end{equation}
\begin{equation}
        C_{I^i g^j} = b_j \int_0^{\chi_\infty} \frac{dN_s^i}{d\chi}\frac{dN_l^j}{d\chi}P_{I m}\left(\frac{\ell}{\chi},\chi\right)\,.
\end{equation}

Each of the 3D power spectra can be expressed in terms of the redshift-dependent amplitude $A_1(z)$, enabling IA modelling to be folded into the same parameter inference framework as lensing and clustering -- see \citep{Blazek_2019} for a detailed theoretical description. In this work we choose to use the NLA model for IA over other more complex modelling frameworks (e.g. Tidal Alignment - Tidal Torquing \cite{Blazek_2019}, the halo model for IA \cite{Fortuna_2020}, or Effective Field Theory of IA \cite{Vlah_2020}). This is because we are primarily focused on the impact of the photo-$z$ model and particularly the inclusion of an outlier population. When introducing this additional complexity in the photo-$z$ space, we benefit from employing a minimal realistic IA model to avoid over-expanding complexity in multiple domains simultaneously and further inflating the parameter space dimensionality. Future work considering outlier rates in combination with more complex IA models would be of value.

\subsection{Parameter Inference}
\label{section:inference}

The cosmological inference process involves sampling from a posterior distribution, which combines prior knowledge (expressed as probability distributions) with a likelihood function that quantifies how well a given parametric model fits the observed data.
Under a typical Gaussian likelihood approximation the natural logarithm of the likelihood $\mathcal{L}$ is given by 
\begin{equation}
    \log \mathcal{L} \propto -0.5 (\mathcal{D}-\mathcal{M}_\theta)^T \Sigma^{-1} (\mathcal{D}-\mathcal{M}_\theta)
\label{equ:like}
\end{equation}
where in a typical real analysis, $\mathcal{D}$ would be a measured datavector. For us, in the simulated analyses of this work, it is a synthetic datavector produced with the fiducial parameters given in Table \ref{tab:fiducial}, acting in place of our observed data. $\mathcal{M}_\theta$ is the datavector produced by our model for a parameter set $\theta$, and $\Sigma$ is the data covariance matrix, assumed to be parameter-independent. In this section we describe the construction of each of these components as used in this analysis. 

To sample efficiently over the parameter space, we employ the Affine Invariant Markov Chain Monte Carlo (MCMC) method implemented using the \texttt{emcee} package \cite{2010CAMCS...5...65G}. This method is well-suited to exploring high-dimensional parameter spaces, as it adapts to the geometry of the posterior distribution and mitigates issues related to correlated parameters.

The full parameter space $\theta$ consists of 7 $w_0 w_a$CDM parameters together with 5 linear galaxy bias parameters $\{b_i\}$, 2 intrinsic alignment parameters for the NLA model $\{a, \eta\}$, and 10 photo-$z$ uncertainty parameters $\{\Delta z^{i}_l, \Delta z^{i}_s\}$. 

\textit{Posterior Predictive Checks:} To assess the suitability of a given model for our data, we employ posterior predictive checks (PPCs), a standard technique in Bayesian model evaluation \cite{gelman_1996}. In the PPC framework, we first draw $N$ parameter vectors $\theta^i$ from the posterior distribution $p(\theta|\mathcal{D})$. For each draw, we generate a replicated datavector $\mathcal{M}_\theta^i$ from the likelihood $p(\mathcal{M}_\theta|\theta^i)$. $p(\theta|\mathcal{D})$ and $p(\mathcal{M}_\theta|\theta)$ together define the posterior predictive distribution:
\begin{equation}
p(\mathcal{M}_\theta|\mathcal{D}) = \int p(\mathcal{M}_\theta|\theta) p(\theta|\mathcal{D}) d\theta.
\end{equation}

This produces a collection of $N$ replicated datavectors, $\mathcal{M}_\theta^i$, that can be directly compared to the observed data $\mathcal{D}$. In this work, we perform a visual comparison by overplotting the model predictions and mock data to identify potential model misspecification, an example of which is shown in Figure~\ref{fig:ppc_s5}.

In addition, we implement the goodness-of-fit test proposed in \cite{gelman_1996} (Equation 3). For this purpose, we use the Mahalanobis distance \cite{mahalanobis1936generalized}, which we define as
\begin{equation}
\chi^2 = (y - \overline{\mathcal{M}}_\theta)^T \Sigma^{-1} (y - \overline{\mathcal{M}}_\theta),
\label{eqn:chi2}
\end{equation}
where $y$ is the datavector which we are testing the goodness-of-fit of and $\overline{\mathcal{M}}_\theta$ is the mean of the posterior predictive datavectors $\mathcal{M}_\theta$. This metric provides a quantitative measure of how well the model predictions match the observed data. If the model is a good representation of the data, we would expect the $\chi^2$ value of the observed (or mock) data to lie within the distribution of $\chi^2$ values computed from the replicated datavectors $\mathcal{M}_\theta^i$.

\textit{Composite Likelihoods:} In this work, we investigate whether leveraging composite likelihoods, following \citep{rau_complike}, can help reduce the effects of model misspecification. A comprehensive review of these methods is provided by \citep{complikereview}. During inference we usually resolve the likelihood given in equation \ref{equ:like} to determine the probability of the proposed parameter set describing the data. After identifying components of the likelihood that are poorly modelled due to misspecification using PPCs, we can isolate these components and down-weight them, arriving at the composite likelihood
\begin{equation}
    \log \mathcal{L} \propto \gamma \log \mathcal{L}_P + \log \mathcal{L}_R\,,
\end{equation}
where subscripts $P$ and $R$ refer to the punitive and representative components, respectively and $\gamma\in [0,1]$ is a weighting factor that limits the importance of the punitive part. To obtain the respective parts $\mathcal L_{P/R}$, we split the data vector $\mathcal{D}$ and the associated covariance matrix $\Sigma$ into blocks 
\begin{equation}
    \mathcal{D}=\begin{pmatrix}
    \mathcal{D}_R\\
    \mathcal{D}_P
    \end{pmatrix}\,,\quad
    \Sigma=\begin{pmatrix}
        \Sigma_{RR} & \Sigma_{RP} \\
    \Sigma_{PR}&\Sigma_{PP}
    \end{pmatrix}\,.
\end{equation}
The likelihood depends on the inverse of the data covariance, which we obtain by considering the covariance as a 2$\times$2 block matrix to identify the separate parts. It is useful to define the two objects 
\begin{equation}\overline\mu=\mathcal{M}_P+\Sigma_{PR}\Sigma_{RR}^{-1}(\mathcal{D}_R-\mathcal{M}_R)\,,\quad
    \overline\Sigma = \Sigma_{PP} - \Sigma_{PR}\Sigma_{RR}^{-1}\Sigma_{RP}\,,
\end{equation}
where $\mathcal{D}$ and $\mathcal{M}$ indicate the mock data and model datavectors respectively. Here $\overline\mu$ and $\overline\Sigma$ are the mean and covariance of the conditional distribution 
\begin{equation}
    P(\mathcal{D}_P|\mathcal{M}_P,\Sigma_{PP},\mathcal{M}_R, \Sigma_{RP}, \mathcal{D}_R) = \mathcal{N}(\overline\mu,\overline\Sigma). 
\end{equation}

We can then extract the punitive component given by the conditional Gaussian distribution of $\mathcal{D}_P$ given $\mathcal{D}_R$
\begin{equation}
    \log \mathcal{L}_{P} \propto - 0.5 \left(\mathcal{D}_P-\overline\mu\right)^T \overline\Sigma^{-1} \left(\mathcal{D}_P-\overline\mu\right),
\end{equation}
and the representative component
\begin{equation}
    \log \mathcal{L}_{R} \propto - 0.5 \left(\mathcal{D}_R-\mathcal{M}_R\right)^T \Sigma_{RR}^{-1} \left(\mathcal{D}_R-\mathcal{M}_R\right)\,.
\end{equation}

We can then repeat the inference process, instead resolving this composite likelihood, with identifiably misspecified components given less importance. Our choice of $\gamma$ can be informed by the PPC from the original inference. As MCMC methods only require the relative probability between samples to evaluate the posteriors, it is not required that $\log \mathcal{L} = \gamma \log \mathcal{L}_{P} + \log \mathcal{L}_R$, but only that the change in probability between samples is conserved. Therefore we do not require the down-weighting of $\log \mathcal{L}_P$ to be counter-weighted in $\log \mathcal{L}_R$.

\section{Results}
\label{section:results}

We performed this analysis in an LSST Year 1 context, which has an expected survey area of $12.3\times 10^3$ deg$^2$ \citep{Leonard_2024}. The fiducial values for the linear galaxy biases $b_i$, the galaxy-galaxy lensing bin combinations, and the scale cuts in $\ell$, as well as the source and lens redshift distributions (Figure \ref{fig:gal_dist}) and the non-Gaussian data covariance matrix, were taken from the LSST DESC SRD \cite{thelsstdarkenergysciencecollaboration2021lsstdarkenergyscience}. This data covariance matrix was also used to produce the noise for scenarios where noise was included in the mock data. 

We used the fiducial parameters shown in Table \ref{tab:fiducial} to generate the mock data used throughout our analysis. In this table we also present the priors placed on these parameters during inference for most scenarios considered in this work and provide the sources that informed these choices. However, when ``centred priors'' are used during inference, all Gaussian priors are defined such that their mean lies at the fiducial value for that parameter. This only affects $w_a$ and $\Delta z^i$ priors when implemented, as all other Gaussian priors are always centred on the fiducial parameters. The uniform priors applied to $b_g^i$ remain unchanged throughout.

\begin{table}[]
    \begin{tabular}{c|c|c|c|c|c|c|c|c|c}
        $\theta$ & $\Omega_M$& $\sigma_8$ & $h$ & $\Omega_B$ &  $n_s$  & $w_0$ & $w_a$ & $a$ & $\eta$ \\\hline
        $\theta_{\rm fid}$ & 0.3156 & 0.830 & 0.6727 & 0.0492 & 0.9645 & -1.00 & 0.05 & 0.44 & -0.7\\
        $\mu_p$ & 0.3156, & 0.831 & 0.6727 & 0.0492 & 0.9645 & -1.00 & 0.00 & 0.44 & -0.7\\
         $\sigma_p$ & \multicolumn{7}{c|}{$\mathbf{\Sigma_p}$ \cite{thelsstdarkenergysciencecollaboration2021lsstdarkenergyscience}} & 0.38 \cite{Abbott_2018} & 2.2 \cite{Abbott_2018} \\\multicolumn{10}{c}{}\\

    $\theta$ & \multicolumn{3}{c|}{$b_g^i$} & \multicolumn{3}{c|}{$\Delta z_l^i$ [$10^{-3}$]} & \multicolumn{3}{c}{$\Delta z_s^i$ [$10^{-3}$]} \\\hline
     $\theta_{\rm fid}$ &
     \multicolumn{3}{c|}{$\{1.56,  1.73, 1.91, 2.10, 2.29 \}$}
     & \multicolumn{3}{c|}{$\{7.1, 1.6, 3.9, 9.0, 7.5 \}$} & 
     \multicolumn{3}{c}{$\{-3.9, 3.8, -0.6, 0.4, 1.6\}$}\\
     prior & 
     \multicolumn{3}{c|}{$\mathcal{U}(0.8,3)$ \cite{Secco_2022,Abbott_2018}} &
     \multicolumn{6}{c}{$\mathcal{N}(0,0.004)$ \cite{Rau_2023}}\\\multicolumn{10}{c}{}\\
    \end{tabular}
    \caption{Fiducial parameters used for the mock data in the MCMC analysis, along with the priors applied during inference and their source. Priors in the upper table are Gaussians with a mean $\mu_p$ and standard deviation $\sigma_p$. The first seven cosmological parameters are drawn from a joint multivariate Gaussian prior
$\mathcal{N}(\mu_p, \mathbf{\Sigma_p})$, where $\mathbf{\Sigma_p}$ corresponds to the Stage III prior parameter covariance matrix including $w_0,w_a$ from \cite{thelsstdarkenergysciencecollaboration2021lsstdarkenergyscience}.}
    \label{tab:fiducial}
\end{table}

\begin{table}[]
    \centering
    \begin{tabular}{c|c|c|c|c}
         \textbf{Scenario} & $\mathbf{f_D}$ & \textbf{Noise} & \textbf{Centred Priors} & \textbf{Composite $\mathcal{L}$}\\ \hline
         1 & 0.00 & No & Yes & No \\
         2 & 0.00 & No & No & No \\
         3 & 0.02 & No & No & No \\
         4 & 0.02 & Yes & No & No \\
         5 & 0.05 & No & No & No \\
         6 & 0.05 & No & No & Yes\\
    \end{tabular}
    \caption{Summary of scenarios considered in this work. $\mathit{f_D}:$ the outlier fraction included in the datavector. \textit{Noise:} Whether Gaussian noise was included in the mock data (Yes) or not (No). \textit{Centred Priors:} Whether all priors were centred on fiducial parameter values (Yes) or were as described in Table \ref{tab:fiducial} (No). \textit{Composite $\mathcal{L}$:} Whether the composite likelihood method described in Section \ref{section:inference} was applied during inference (Yes) or not (No).}
    \label{tab:scenarios}  
\end{table}

In this work, we consider six scenarios summarised in Table \ref{tab:scenarios}:
\begin{enumerate}
    \item We produce mock data using the fiducial parameters from Table \ref{tab:fiducial}, and include no outlier fraction and no noise. We use centred priors, modifying the priors applied such that all Gaussian priors are centred on the fiducial parameters used to produce the data. We then carry out the inference process using the model described in Section \ref{section:modelling}. This could be considered a perfectly modelled case, and we use this as a reference for the other scenarios we consider.
    \item We repeat the inference from Scenario 1, but implement the priors as described in Table \ref{tab:fiducial}. This simulates a standard inference process where all systematics are modelled, and there is no outlier fraction present in the data.
    \item We add an outlier fraction of $f_D$ = 0.02 to the mock data. As a parametrisation of the outlier fraction is not included in the model, this allows us to determine the robustness of the inference process in the case of a small misspecification error in outlier fraction.
    \item We include an outlier fraction of $f_D=0.02$ in the mock data and add Gaussian noise to determine the effect of a small outlier fraction on parameter inference in a more realistic case.
    \item We include an outlier fraction of $f_D=0.05$ in the mock data to determine the effect of an increased outlier fraction on parameter inference.
    \item We include an outlier fraction of $f_D=0.05$ in the mock data and down-weight the likelihood components impacted by the outlier fraction by a factor of $\gamma=0.05$ using the method described in Section \ref{section:inference}. We use this to test whether this method can improve parameter constraints in the presence of misspecification.
\end{enumerate}

We confirmed that MCMC chains were converged by calculating the Gelman-Rubin statistic \citep{Gelman_Rubin_1992} between the first half and second half of the chain along iterations, after removing burn-in. 

\subsection{Inferred Cosmology Response to Outlier Misspecification}
\label{section:cosmo_results}
\begin{figure}
    \centering
    \includegraphics[width=\linewidth]{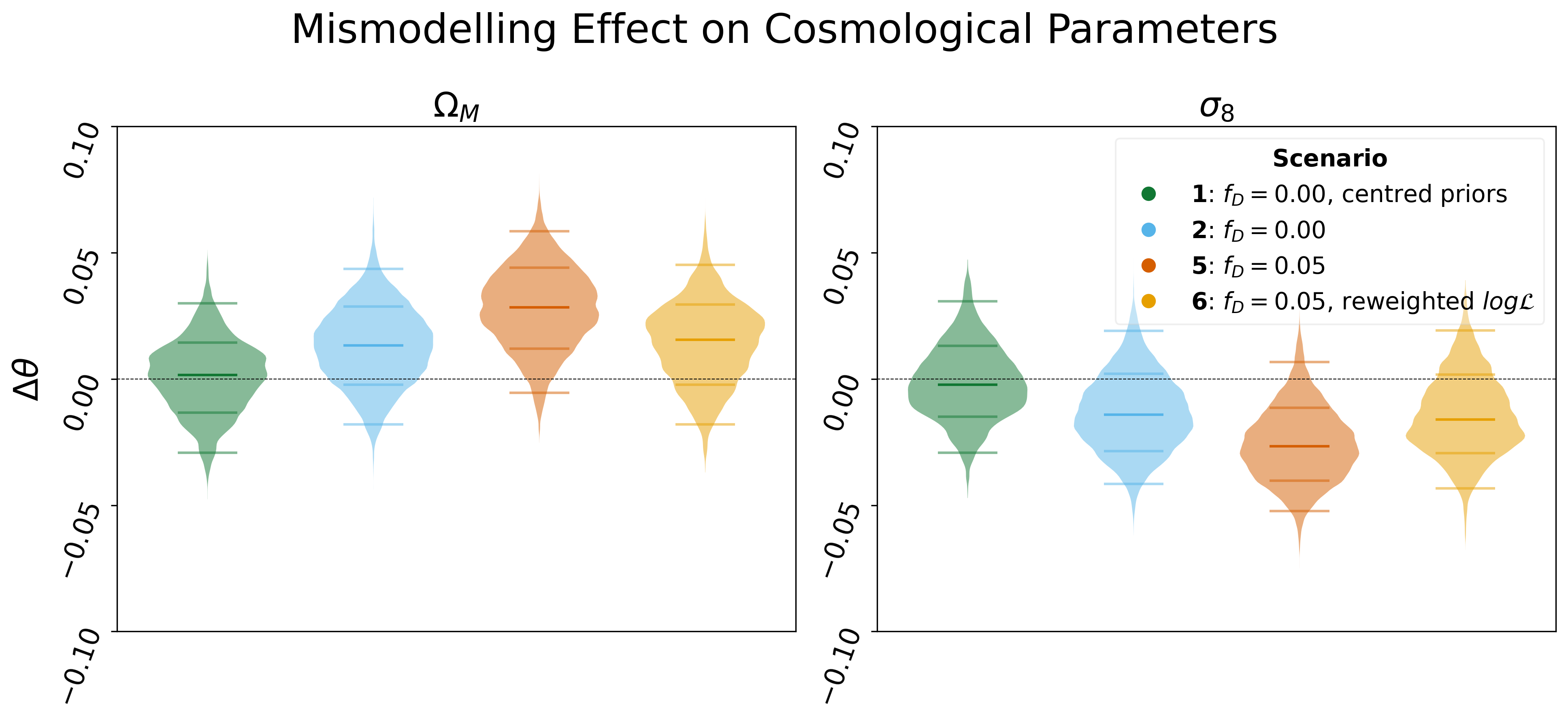}
    \qquad
    \includegraphics[width=\linewidth]{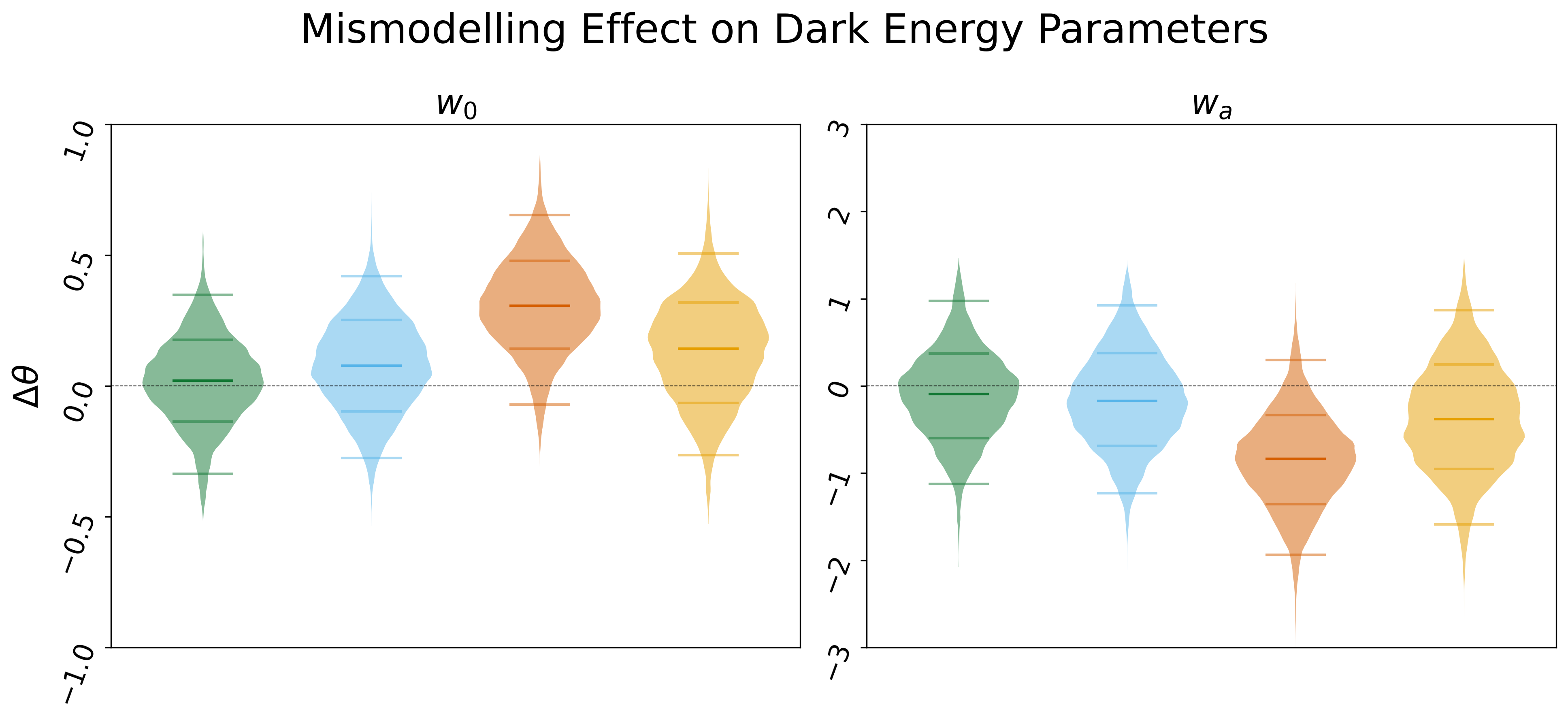}
    \caption{The posterior probability distributions for cosmological parameters $\Omega_M, \sigma_8$ (upper) and dark energy parameters (lower) for a subset of the scenarios described in Table \ref{tab:scenarios}. $\Delta \theta$ indicates the difference between the maximum a posteriori and the fiducial value (Table \ref{tab:fiducial}), where $\Delta \theta = 0 $ is indicated with the dashed black line. Mean and 1$\sigma$ and 2$\sigma$ are indicated by the dark and light lines.}
    \label{fig:violins}
\end{figure}

We show the results of the parameter inference process for the cosmological and dark energy parameters in Figure \ref{fig:violins}, focussing here on the most illustrative of the Scenarios from Table \ref{tab:scenarios} (with results for all Scenarios shown in Appendix \ref{app:full}). We found that we were able to obtain cosmological parameters ($\Omega_M$ and $\sigma_8$) near the fiducial values when we included no outliers in the data and centred all priors on the fiducial values used to build the datavector (Scenario 1), which we can consider a perfectly modelled scenario. A bias of $1\sigma$ in the cosmological parameters was produced by applying priors which are not centred on the true parameter values as described in Table \ref{tab:fiducial} (Scenario 2), with a milder effect on the dark energy parameters. Once an unmodelled outlier fraction of $f_D=0.05$ was included in the data (Scenario 5), a bias of $1.8\sigma$ was produced in $\Omega_M, \sigma_8,$ and $w_0$ with a slightly smaller bias produced in $w_a$.

\begin{figure}
    \centering
    \includegraphics[width=0.8\linewidth]{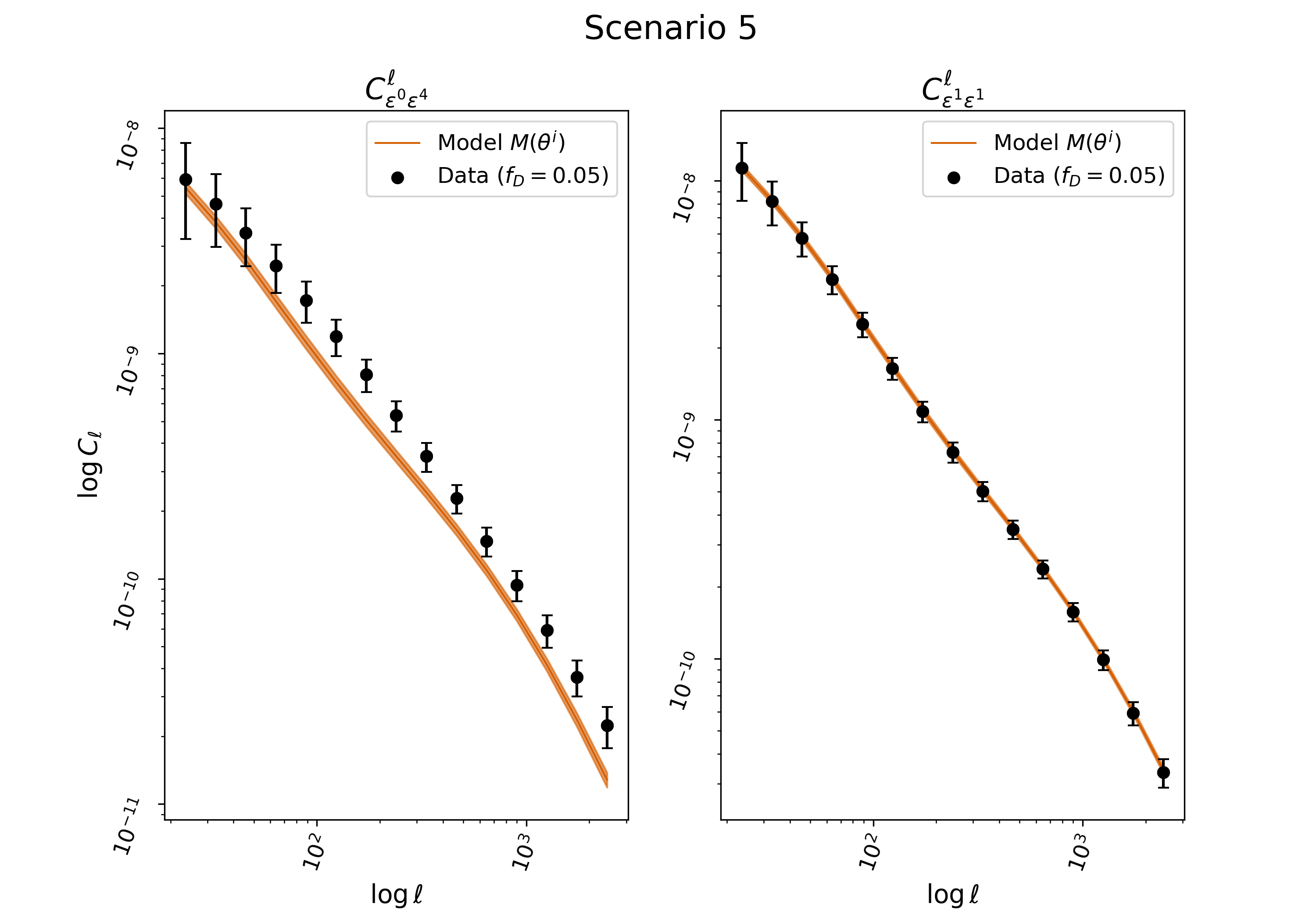}
    \caption{The posterior predictive diagnostic plots for Scenario 5 for two weak lensing angular power spectra in the mock data. Catastrophic outliers included in the lowest bin affect cross-correlations significantly (left) while leaving the autocorrelation in higher bins unaffected (right). We plot the model as the mean of the 500 datavectors produced from sampling the posterior, with the shaded region indicating the 2$\sigma$ error, and plot the mock data as data points with 2$\sigma$ error bars drawn from the data covariance matrix.}
    \label{fig:ppc_s5}
\end{figure}

\begin{figure}
    \centering
    \includegraphics[width=\linewidth]{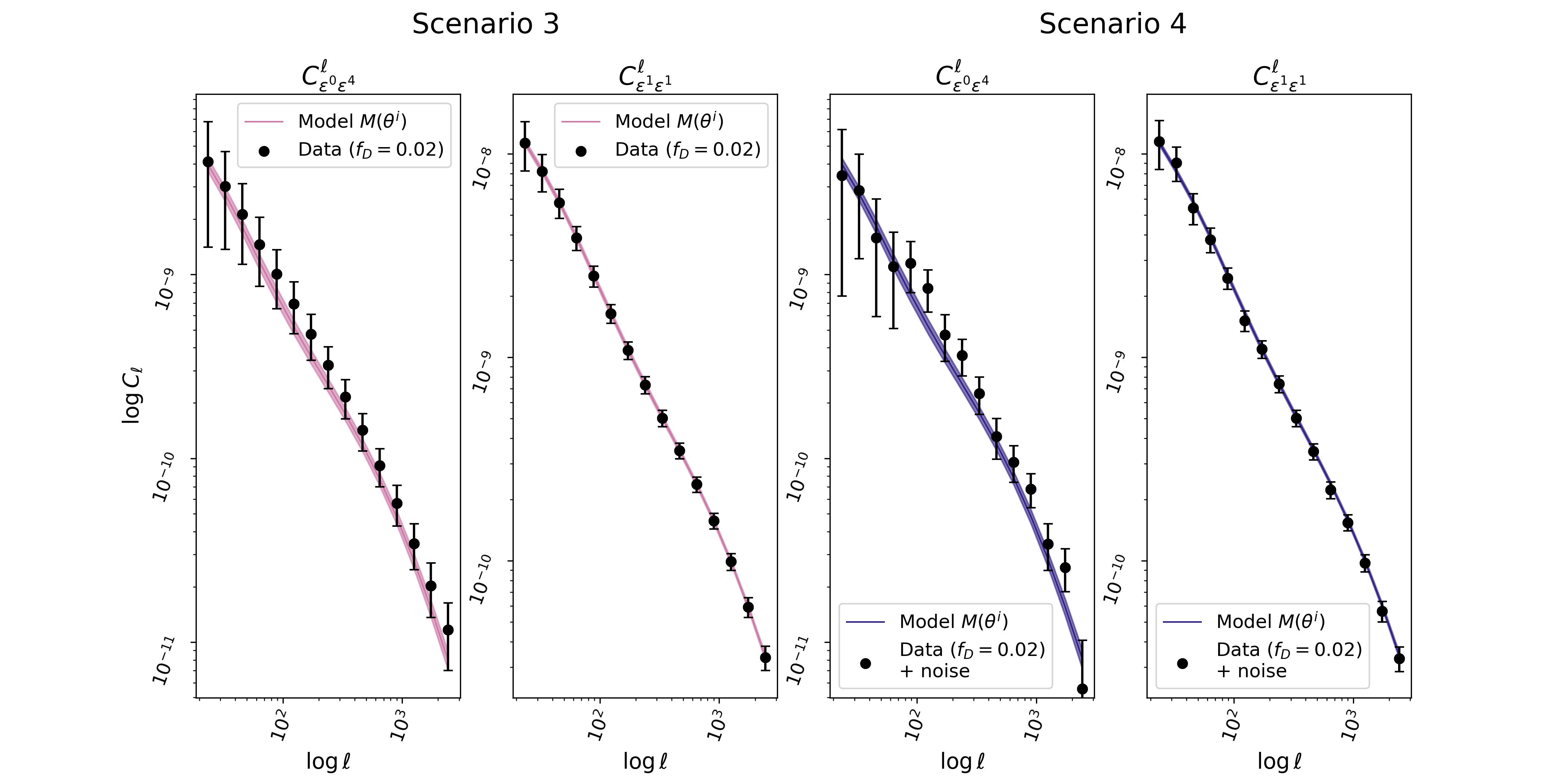}
    \caption{The posterior predictive diagnostic plots for Scenario 3 and 4 for two weak lensing angular power spectra in the mock data. With an outlier fraction of $f=0.02$ a bias is produced (left) that is not present in model components that don't contain the outlier population (centre left). However, this bias can be obscured when noise is added to the mock data (centre right). Nevertheless, one can still see disagreement between the replicated data and the mock data when comparing to components of the model not affected by the outlier fraction (right). We plot the model as the mean of the 500 datavectors produced from sampling the posterior, with the shaded region indicating the 2$\sigma$ error, and plot the mock data as data points with 2$\sigma$ error bars drawn from the data covariance matrix.}
    \label{fig:ppc_s3n4}
\end{figure}

We can detect the poor fit of the model to the data visually by implementing PPCs, as shown in Figure \ref{fig:ppc_s5}. Considering Scenario 5, we plot the mock data for the cosmic shear cross-correlation $C^\ell_{\epsilon^0\epsilon^4}$ along with the mean and standard deviation of the reconstructed data from 500 samples from the posterior (from here on referred to as replicated data). We also plot this for $C^\ell_{\epsilon^1\epsilon^1}$; a component of the datavector that is not directly affected by misspecification in the outlier fraction. The outlier population induces a bias in the data that the model is unable to fit to, producing a poor reproduction of the angular power spectra. This misspecification only produces a poor fit in angular power spectra that include the tomographic redshift bin $\frac{dN^0_s}{dz}$ containing the outlier population (i.e. $C_{\epsilon^0 \epsilon ^{0-4}}^\ell$ ), while spectra for other bin combinations (e.g. $C_{\epsilon^1 \epsilon^1}^\ell$) remain unaffected. By visual comparison then, we can identify cases of misspecification and inform where in the model the misspecification may be implemented. 

We draw attention to the difficulty of identifying this misspecification when noise is included on the mock data in Figure \ref{fig:ppc_s3n4}, where we display the mock data and replicated data for Scenarios 3 and 4. These have identical set-ups apart from the inclusion of Gaussian noise in the mock data in Scenario 4. In the noiseless case of Scenario 3, a consistent bias is still clear for an outlier fraction of $f=0.02$, even though the replicated data are within $2\sigma$ of the mock data. In the noisy case, there is a clear divergence between the mock and replicated data, with the mock data biased by over $2\sigma$ at some values of $\ell$. However, there is not a consistent positive or negative bias between the mock data and the replicated data in the presence of noise. Therefore, it may be difficult to identify cases of misspecification when analysing noisy data. 

\begin{figure}
    \centering
    \includegraphics[width=\linewidth]{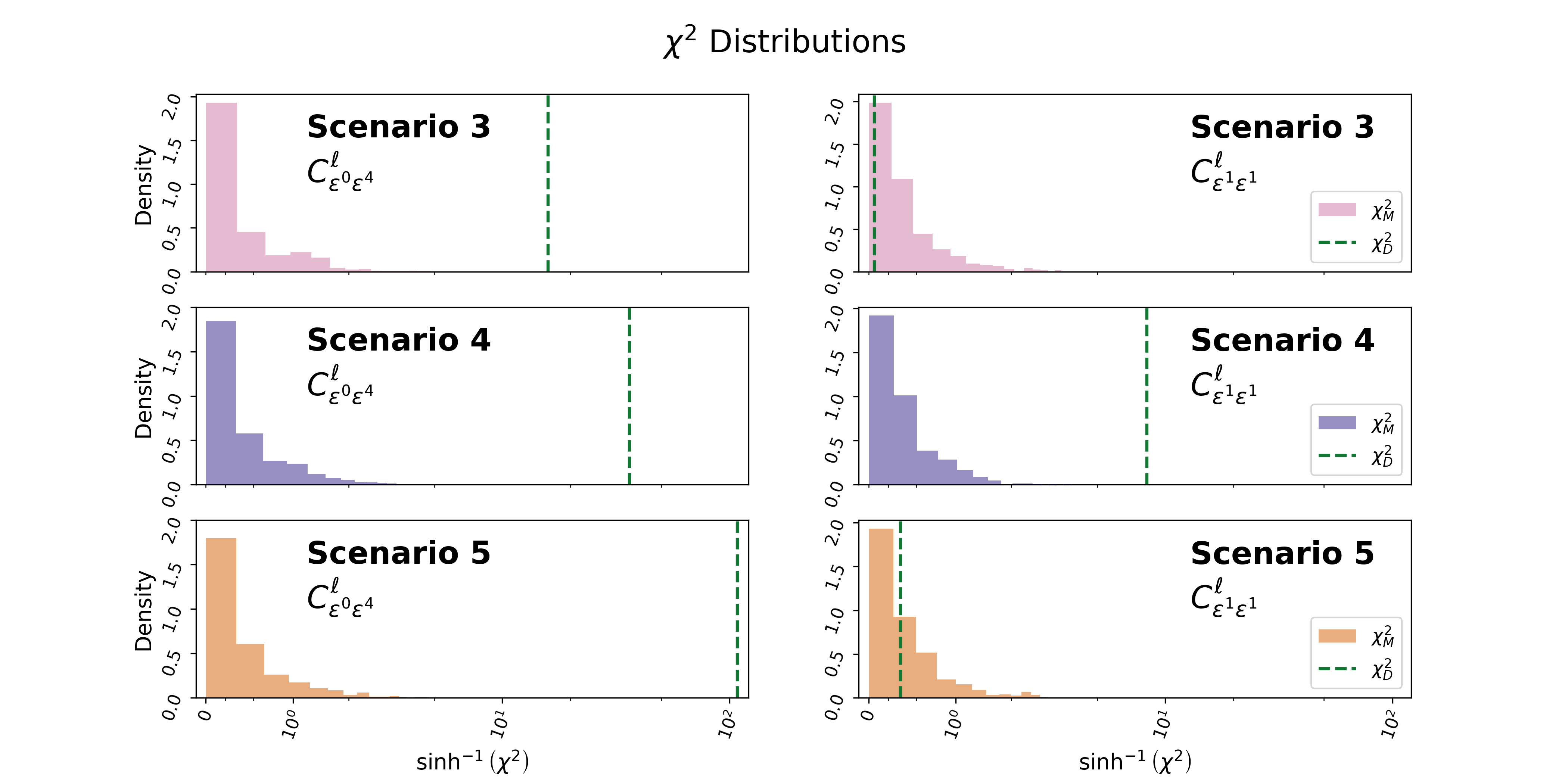}
    \caption{The distributions of $\chi^2_{M}$ for the 500 replicated datavectors for Scenarios 3, 4, and 5 for a component of the datavector where an outlier fraction is present in one of the cross-correlated redshift bins (left) vs. not present (right), with the $\chi^2_{D}$ value in the mock data shown as a green dashed line. The x-axis is rescaled as $\sinh^{-1}(\chi^2)$ to allow for clear visual comparison over the large ranges considered.}
    \label{fig:chi2}
\end{figure}

We implement Equation \ref{eqn:chi2} to determine $\chi^2_{M^i}$ for each of the 500 replicated datavectors, as well as $\chi_D^2$ for the mock data, to determine whether the model is representative of the data. We plot the results of this test in Figure \ref{fig:chi2}. We find that in Scenarios 3 and 5, it is clear that $C^\ell_{\epsilon^0\epsilon^4}$ is misspecified while for $C^\ell_{\epsilon^1\epsilon^1}$ the model is representative of the data. This is less clear when we consider Scenario 4, where the noise causes both $C^\ell$s to be indicative of a poor fit. However, the $C^\ell_{\epsilon^0\epsilon^4}$ component is a significantly worse fit when compared to $C^\ell_{\epsilon^1\epsilon^1}$. Therefore, when considering the relative goodness-of-fit of components of the data it is possible to identify poorly modelled components of the data even in the presence of noise. 

Now that the components of the datavector impacted by the misspecification have been identified using PPCs, we can choose to down-weight the likelihood of these as discussed in Section \ref{section:inference}. Due to the large discrepancy between the model and data for this section of the datavector, we choose to heavily down-weight the punitive component with $\gamma=0.05$. Once this down-weighting is implemented, the bias in the cosmological and dark energy parameters is reduced to less than $1\sigma$. This shows that the composite likelihood approach is promising in mitigating the impact of unmodelled photo-$z$ outliers.

\subsection{IA and photo-$z$ Parameters}
\label{section:IA_results}
\begin{figure}
    \centering
    \includegraphics[width=\linewidth]{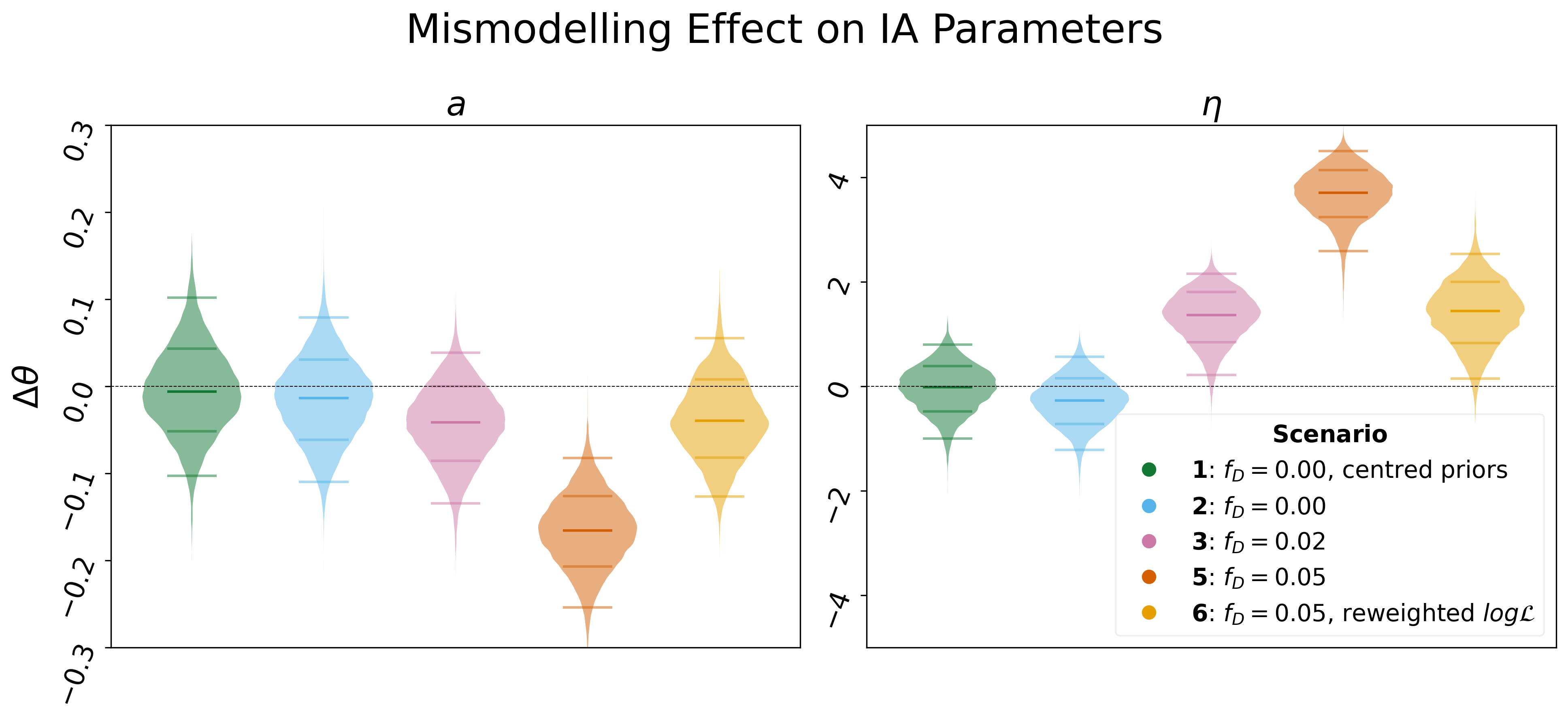}
    \qquad
    \includegraphics[width=\linewidth]{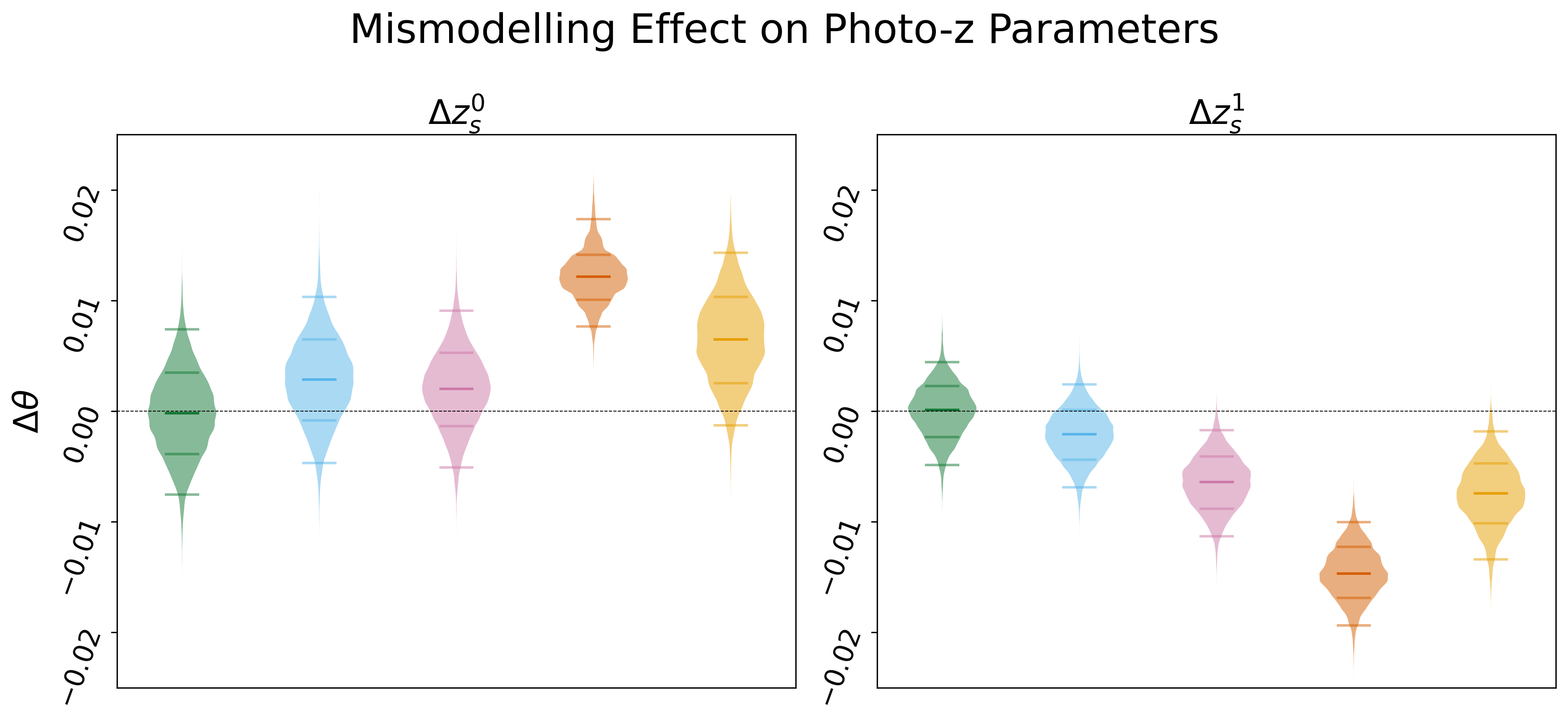}
    \caption{The posterior probability distributions for the NLA IA parameters; the overall amplitude $a$  and the redshift-dependence $\eta$ (upper) and for the $\Delta z_s^i$ parameters for the first two source redshift bins (lower) for the scenarios described in Table \ref{tab:scenarios}. $\Delta \theta$ indicates the difference between the maximum {\it a posteriori} and fiducial value (Table \ref{tab:fiducial}), where $\Delta \theta = 0 $ is indicated with the dashed black line. Mean and 1$\sigma$ and 2$\sigma$ are indicated by the dark and light lines.}
    \label{fig:violin_ia}
\end{figure}

We also visualise the posteriors on the IA parameter and on a subset of photo-$z$ parameters for the scenarios described in Table \ref{tab:scenarios}, in addition to the case of the misspecification of a smaller outlier population, in Figure \ref{fig:violin_ia}. We see that we are able to achieve reasonably unbiased constraints when no outlier rate is included. An outlier fraction of $f_D=0.02$ produces a $1\sigma$ bias in the overall IA amplitude $a$ and a $2\sigma$ bias in the IA redshift-dependence $\eta$. This bias increases to $4\sigma$ in $a$ and $8\sigma$ in $\eta$ when this outlier fraction was increased to $f_D=0.05$. When a reweighting of the likelihood was applied, this bias was reduced to a level comparable to the $f_D=0.02$ case. The presence of the outlier population produces a bias in the $\Delta z_s^i$ parameter, both in the bin containing the population and in the neighbouring bin. We expect that this is due to the high degeneracy between the photo-$z$ uncertainty parameters. We find that the bins further from the one containing the outlier population (not shown) are not significantly biased. We find that the likelihood reweighting is also able to reduce the bias in the $\Delta z_s^i$ parameters. 

\section{Discussion and Conclusions}
\label{section:discuss}

We first note that in considering the results shown in Figures \ref{fig:violins} and \ref{fig:violin_ia}, the only case where the peak of the posterior distribution very closely aligns with the fiducial (true) parameters is that where the prior distributions were artificially centred on those fiducial parameters. We see that in all other cases considered, even with a completely accurate model, constraints are biased, in some parameters to nearly $1\sigma$. This reinforces the fact we must be aware of possible effects on our constraints from informative prior distributions, which in reality cannot be generically centred on the true underlying parameter values describing our data.

In this work, we have considered a somewhat optimistic case where the model can perfectly reproduce the data, there is only one tomographic redshift bin with outliers, and those outliers are localised in one redshift region. This is instead of a more realistic case like that considered in \cite{zhang2025forecastingimpactsourcegalaxy} where all bins contain outlier populations, some with multiple such populations. However, even in the optimistic case considered here, a single outlier population can bias cosmological and dark energy parameters up to almost $2\sigma$ when it is not included in the model. We also see that misspecification of the outlier fraction can significantly bias the IA parameters. This is not a surprising effect, since here we have considered an outlier population in the 0th redshift bin, and it is generally understood that lower-redshift cosmic shear holds more information about IA than the higher-redshift cases.  

In order to avoid these biases, it may be tempting to propose that we simply model these populations during the inference stage. However, this would require an understanding of the distribution of these outlier galaxies that is not currently possible to achieve with the spectroscopic samples available at the depths considered in this work \cite{Newman_2022}. Furthermore, in this work we have considered only a single outlier population in a singular tomographic redshift bin, which only requires one additional model parameter. It is more likely that real data will contain multiple outlier populations across multiple bins, with shapes that are not well understood. To accurately model this would massively expand the parameter space of the analysis, and the degeneracy between these and other nuisance parameters (as shown in Figure \ref{fig:violin_ia}) risks making the likelihood intractably complex. However, as long as this misspecification can be identified through methods like posterior predictive checks, we have shown that the parameter bias can be reduced by down-weighting the poorly modelled components of the likelihood, with minimal effect on the constraining power. 

In this case, with a single outlier population, we are aided somewhat in the application of the composite likelihood method by the simplicity of the scenario. A question which might arise in the case with multiple outlier populations (as in \cite{zhang2025forecastingimpactsourcegalaxy}) is whether it would be possible to effectively apply the composite likelihood method to down-weight relevant components of the data vector in the case where there are outlier populations in multiple redshift bins. Fortunately, the PPC method can help us to identify the most severely affected components of the datavector in this scenario, and an iterative approach can be used to achieve a balance between retaining information and achieving unbiased results. The composite likelihood method is also flexible: the likelihood can in fact be split into more than two components, with different $\gamma$ factors being chosen to differentially down-weight different segments of the data vector that are more or less impacted by misspecification.

One might also be justifiably concerned that down-weighting the portions of the likelihood containing the outlier rate would negatively impact the constraints on the IA and photo-$z$ parameters, due to the importance of lower redshift information for constraining IA and the degeneracy between these parameters. We have found that, promisingly, the constraints on these parameters do not worsen with this method (Figure \ref{fig:violin_ia}). In fact, we find that applying this method reduces the bias in IA parameters of the NLA model to be comparable with a reduction in the outlier fraction. Future applications to more complex or realistic cases will require careful tuning of the weightings of the likelihood, potentially informed by both posterior predictive checks and simulations. 

While we have shown that the poorly fitting components of the model are easily visually identified in the case of noiseless simulated data, we have also demonstrated that this is non-trivial when considering more realistic noisy data (Figure \ref{fig:ppc_s3n4}). We have suggested methods to nevertheless use PPCs and related tools to support the composite likelihood method in this case, however, further work to build up an infrastructure for using composite likelihoods in the case of realistic noisy data would be a valuable step towards applying this method in a real analysis.

We have shown that the composite likelihood approach has promise for mitigating the impact of an outlier population. We further suggest that it has the potential to be a more generally useful tool to mitigate model misspecification. There is scope for further exploration of the applicability of this method and its utility to mitigate issues related to other systematic effects, including e.g. baryonic feedback or other non-linear structure formation processes.

We provide now a clear, point-by-point summary of the main take-aways of this work:

\begin{itemize}
\item Model misspecification for the outlier fraction has the potential to significantly bias cosmological and dark energy parameter constraints, with shifts approaching $2\sigma$.
\item Even when the data is generated from a perfectly known model, miscentred priors can introduce parameter biases up to $1\sigma$.
\item The outlier fraction induces a particularly large bias in the IA parameters, which can reach up to $8\sigma$ for reasonable outlier fractions. 
\item We demonstrated that posterior predictive checks can effectively identify poorly modelled components of the data vector, specifically those stemming from an unmodelled outlier population.
\item Using a composite likelihood approach to down-weight affected components — guided by posterior predictive diagnostics — reduced parameter biases to below $1\sigma$ without degrading constraints on  intrinsic alignment and photometric redshift shifts, despite the down-weighted observables being important for constraining them.
\end{itemize}

These findings highlight the importance of identifying and mitigating model misspecification in cosmological analyses from realistically complex redshift distributions. While our analysis focused on an optimistic and simplified scenario, the techniques developed here can inform future studies dealing with more realistic data. In such cases, tuning the composite likelihood weights—potentially using a combination of posterior predictive checks and simulation-based diagnostics—will be essential. Additionally, further work is needed to assess the robustness of these methods under the presence of noise, which will be a key step toward applying them to real data. It would also be interesting to extend this work to assess the impact of IA and photo-$z$ systematics on beyond 2-point and map-based statistics. Future work could investigate the potential of such techniques to identify systematic effects and break degeneracies, building on e.g. \citep{Friedrich2025, Porqueres2023, Zuercher2023, Halder2023, LSST_Lanzieri2023}. 

\acknowledgments

CMM acknowledges the UK Science and Technology Facilities Council (STFC) for support from grant No. ST/W006790/1. CMM was supported by an STFC-funded Long-Term Attachment ST/W006790/1. MMR acknowledges support by the  IES\textbackslash R1\textbackslash 241442 Royal Society International Exchanges 2024 grant. CU was supported by the STFC Astronomy Theory Consolidated Grant ST/W001020/1 from UK Research \& Innovation and by the European Union (ERC StG, LSS\_BeyondAverage, 101075919). CDL was supported by the Science and Technology Facilities Council (STFC) [grant No. UKRI1172]. This work used the DiRAC Data Intensive service (DIaL2 / DIaL 3) at the University of Leicester, managed by the University of Leicester Research Computing Service on behalf of the STFC DiRAC HPC Facility (www.dirac.ac.uk). The DiRAC service at Leicester was funded by BEIS, UKRI and STFC capital funding and STFC operations grants. DiRAC is part of the UKRI Digital Research Infrastructure. The figures in this work were generated using Matplotlib \citep{Hunter:2007} making use of NumPy \citep{harris2020array} and SciPy \citep{2020SciPy-NMeth}.

Visualisation of the full parameter posterior distributions for all scenarios considered in this work are provided at \url{https://github.com/carolymmmm/3x2pt_contours}.

\bibliographystyle{JHEP}
\bibliography{ref.bib}

\appendix
\section{Full Parameter Constraints}
\label{app:full}

We present the cosmological, dark energy, and IA parameter constraints for each scenario considered in this work in Table \ref{tab:scenario_cosmo}. We also show these in the form of violin plots in Figure \ref{fig:violins_full}. 

\begin{sidewaystable}[ht]
    \centering
    \begin{tabular}{c|c|c|c|c|c|c}
         \textbf{Scenario}& 1 & 2 & 3 & 4 & 5 & 6 \\
         \hline
         $\Omega_M$  & $0.317^{+0.013}_{-0.015}$ & $0.329^{+0.015}_{-0.016}$ & $0.330^{+0.014}_{-0.015}$ & $0.347^{+0.018}_{-0.016}$ & $0.344^{+0.016}_{-0.016}$ & $0.331^{+0.014}_{-0.018}$\\
         $\sigma_8$  & $0.828^{+0.015}_{-0.013}$ & $0.816^{+0.016}_{-0.014}$ & $0.0816^{+0.015}_{-0.014}$ & $0.801^{+0.015}_{-0.015}$ & $0.803^{+0.015}_{-0.014}$ & $0.814^{+0.018}_{-0.013}$\\
         $w_0$  & $-0.979^{+0.155}_{-0.156}$ & $-0.922^{+0.175}_{-0.176}$ & $-0.882^{+0.116}_{-0.175}$ & $-0.683^{+0.179}_{-0.188}$ & $-0.693^{+0.172}_{-0.165}$ & $-0.857^{+0.176}_{-0.207}$\\
         $w_a$   & $-0.041^{+0.461}_{-0.506}$ & $-0.122^{+0.547}_{-0.514}$ & $-0.249^{+0.518}_{-0.509}$ & $-0.837^{+0.569}_{-0.519}$ & $-0.786^{+0.500}_{-0.521}$ & $-0.331^{+0.627}_{-0.572}$\\
         $a$   & $0.434^{+0.049}_{-0.046}$ & $0.427^{+0.044}_{-0.048}$ & $0.399^{+0.041}_{-0.044}$ & $0.389^{+0.041}_{-0.046}$ & $0.274^{+0.040}_{-0.041}$ & $0.401^{+0.047}_{-0.043}$\\
         $\eta$   & $-0.717^{+0.405}_{-0.464}$ & $-0.971^{+0.424}_{-0.446}$ & $0.669^{+0.438}_{-0.523}$ & $0.789^{+0.632}_{-0.640}$ & $3.009^{+0.430}_{-0.470}$ & $0.740^{+0.558}_{-0.608}$\\
         $\Delta z_s^0$  & $-0.004^{+0.004}_{-0.004}$ & $-0.001^{+0.004}_{-0.004}$ & $-0.002^{+0.003}_{-0.003}$ & $-0.004^{+0.004}_{-0.004}$ & $0.008^{+0.002}_{-0.002}$ & $0.003^{+0.004}_{-0.004}$ \\
         $\Delta z_s^1$  & $0.004^{+0.002}_{-0.002}$ & $0.002^{+0.002}_{-0.002}$ & $-0.003^{+0.002}_{-0.002}$ & $-0.005^{+0.002}_{-0.002}$ & $-0.011^{+0.002}_{-0.002}$ & $-0.004^{+0.003}_{-0.003}$\\
    \end{tabular}
    \caption{Cosmological parameter constraints and their $1\sigma$ values for each scenario. Columns correspond to those in Table~\ref{tab:scenarios}.}
    \label{tab:scenario_cosmo}
\end{sidewaystable}

\begin{figure}
    \centering
    \includegraphics[width=\linewidth]{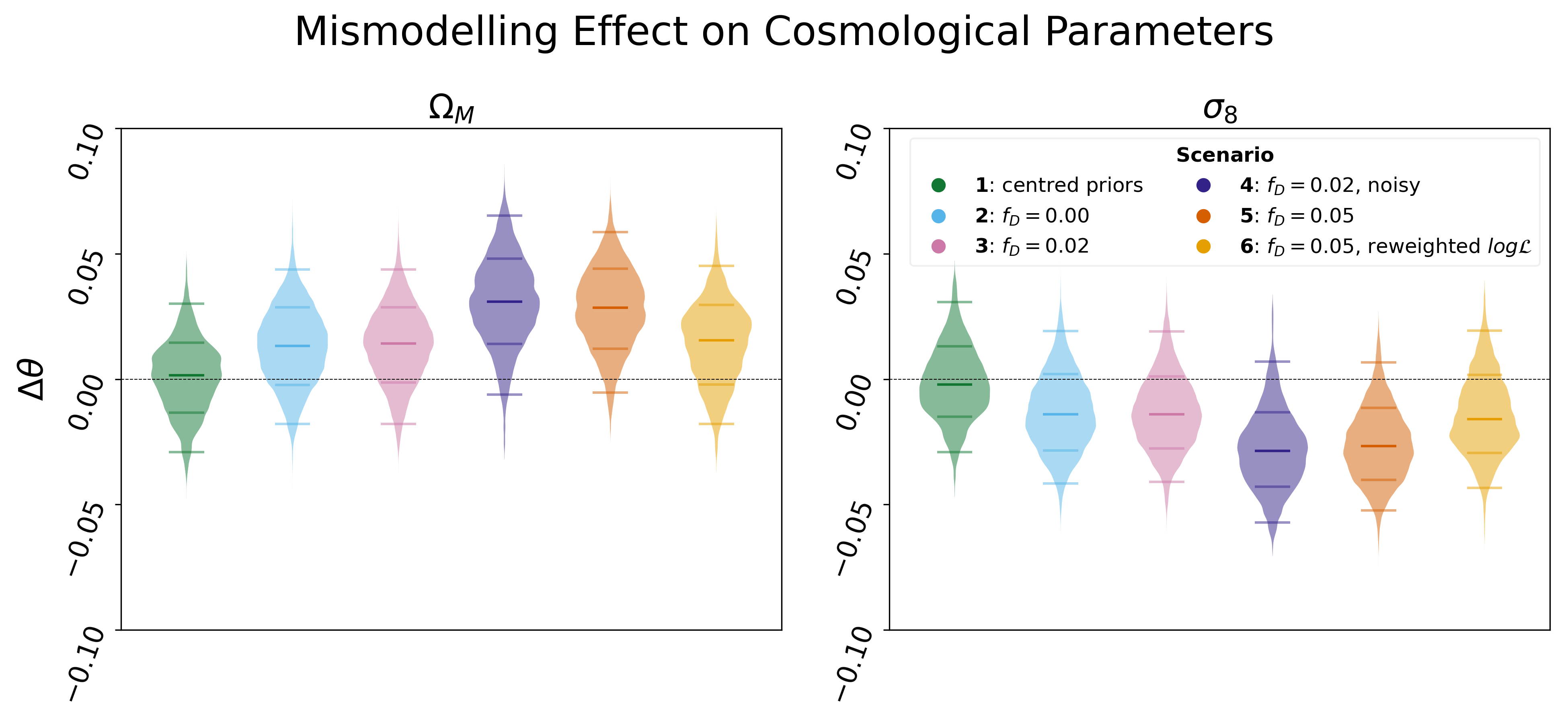}
    \qquad
    \includegraphics[width=\linewidth]{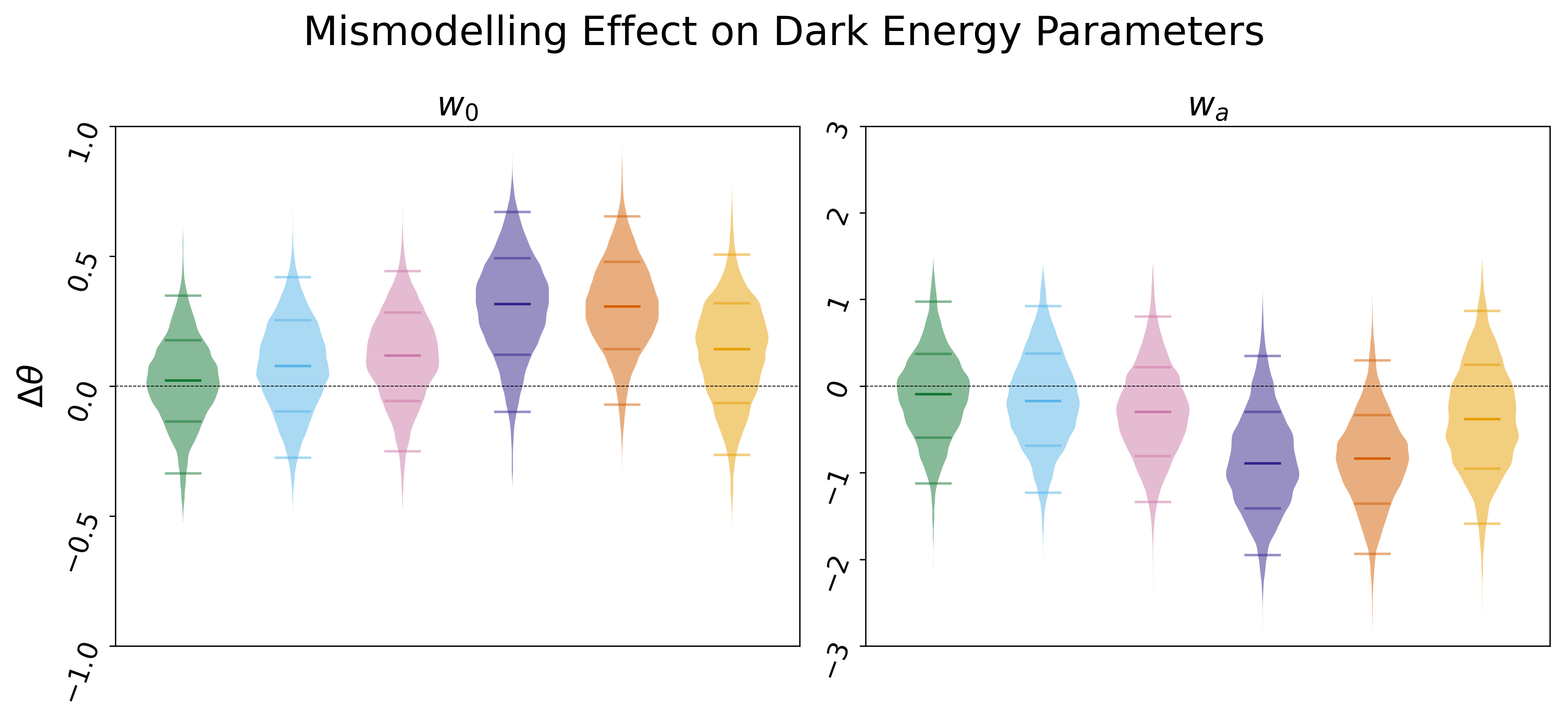}
    \caption{The posterior probability distributions for cosmological parameters $\Omega_M, \sigma_8$ (upper) and dark energy parameters (lower) for a range of scenarios described in Table \ref{tab:scenarios}. $\Delta \theta$ indicates the difference between the maximum a posteriori and the fiducial value (Table \ref{tab:fiducial}), where $\Delta \theta = 0 $ is indicated with the dashed black line. Mean and 1$\sigma$ and 2$\sigma$ are indicated by the dark and light lines.}
    \label{fig:violins_full}
\end{figure}

\begin{figure}
    \centering
    \includegraphics[width=\linewidth]{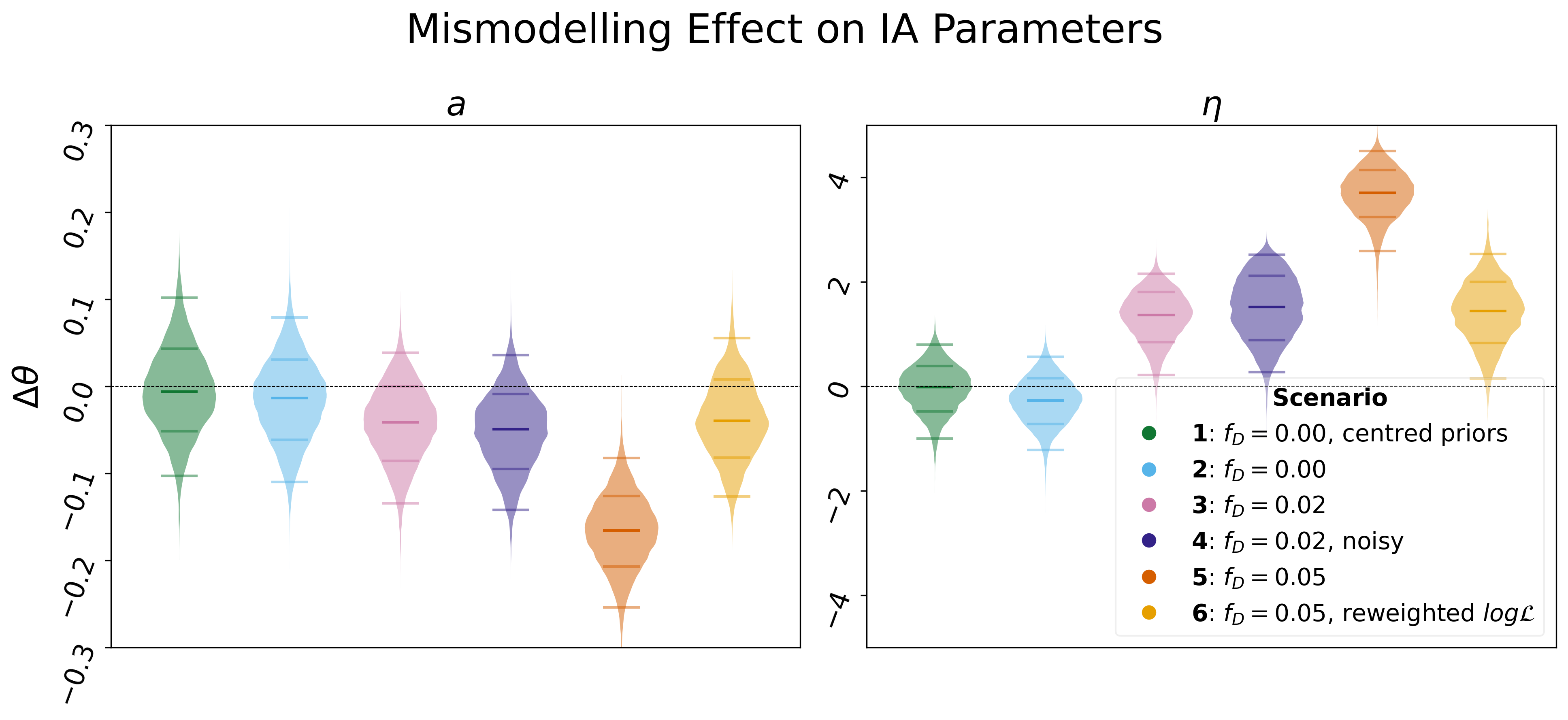}
    \qquad
    \includegraphics[width=\linewidth]{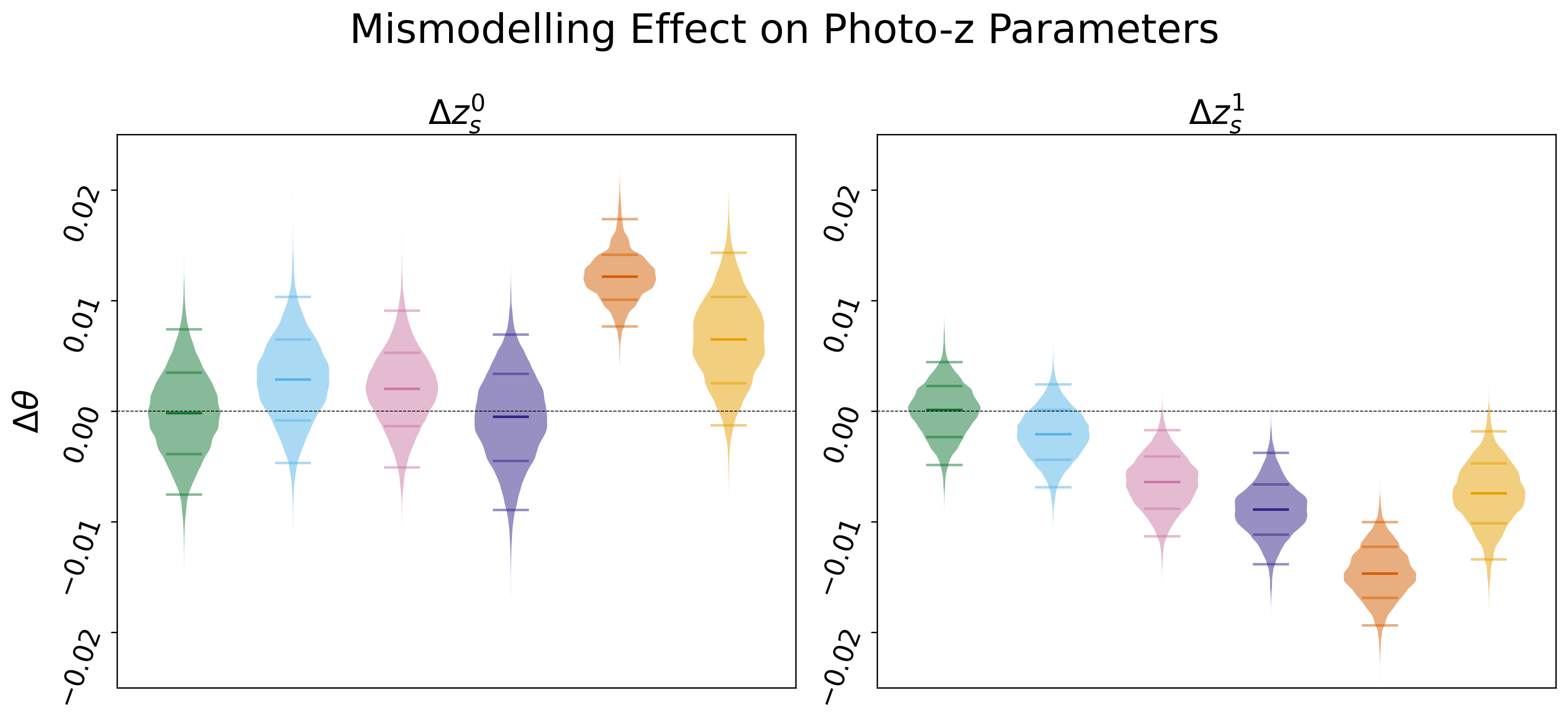}
    \caption{The posterior probability distributions for IA parameters and photo-$z$ uncertainty parameters (lower) for a range of scenarios described in Table \ref{tab:scenarios}. $\Delta \theta$ indicates the difference between the maximum a posteriori and the fiducial value (Table \ref{tab:fiducial}), where $\Delta \theta = 0 $ is indicated with the dashed black line. Mean and 1$\sigma$ and 2$\sigma$ are indicated by the dark and light lines.}
    \label{fig:violins_full}
\end{figure}

\end{document}